\documentclass[11pt,preprint]{aastex}
\pdfoutput=1
\usepackage{graphicx}

\let\oldfootsep=\footnotesep
\newcommand\ltsima{$\; \buildrel <\over\sim \;$}
\newcommand\simlt{\lower.5ex\hbox{\ltsima}}
\newcommand\gtsima{$\; \buildrel >\over\sim \;$}
\newcommand\simgt{\lower.5ex\hbox{\gtsima}}
\newcommand\etal{et~al.}

\newcommand\msun {M_\odot}

\newcommand\mearth {{M_\oplus}}

\newcommand\pac{Paczy{\'n}ski }

\newcommand\rep {\tilde{r}_E}

\shorttitle{}
\shortauthors{Bennett et al}

\begin{document}


\title{The First Circumbinary Planet Found by Microlensing: OGLE-2007-BLG-349L(AB)c}


\author{D.P.~Bennett\altaffilmark{1,2,M,P},
S.H.~Rhie\altaffilmark{\dagger,2},
A.~Udalski$^{3,O}$,
A.~Gould\altaffilmark{4,5,6,\mu},
Y.~Tsapras$^{7,8,R}$,
D.~Kubas\altaffilmark{9,P},
I.A.~Bond\altaffilmark{10,M}, 
J.~Greenhill\altaffilmark{\dagger,11,P},
A.~Cassan$^{9,P}$,
N.J.~Rattenbury$^{12,M}$, 
T.S. Boyajian\altaffilmark{13},
J.~Luhn\altaffilmark{14},
M.T.~Penny\altaffilmark{4},
J.~Anderson\altaffilmark{15},
\\
and \\
F.~Abe$^{16}$, 
A.~Bhattacharya\altaffilmark{2},
C.S.~Botzler$^{12}$, 
M.~Donachie$^{12}$,
M.~Freeman$^{12}$, 
A.~Fukui$^{17}$, 
Y.~Hirao$^{18}$, 
Y.~Itow$^{16}$,  
N.~Koshimoto$^{18}$,
M.C.A.~Li$^{12}$,
C.H.~Ling$^{10}$, 
K.~Masuda$^{16}$,  
Y.~Matsubara$^{16}$, 
Y.~Muraki$^{16}$, 
M.~Nagakane$^{18}$,
K.~Ohnishi$^{19}$, 
H.~Oyokawa$^{16}$, 
Y.C.~Perrott$^{12}$,
To.~Saito$^{20}$,
A.~Sharan$^{12}$,
D.J.~Sullivan$^{21}$, 
T.~Sumi$^{18}$,
D.~Suzuki$^{1,2}$,
P.J.~Tristram$^{22}$,
A. Yonehara$^{23}$,
P.C.M.~Yock$^{12}$, \\ (The MOA Collaboration)\\
M.K.~Szyma{\'n}ski$^3$,
I.~Soszy{\'n}ski$^3$,
K.~Ulaczyk$^3$,
{\L}.~Wyrzykowski$^3$, \\ (The OGLE Collaboration)\\
W.~Allen$^{24}$,
D.~DePoy$^{25}$,
A.~Gal-Yam$^{26}$,
B.S.~Gaudi$^{4}$,
C.~Han$^{27}$,
I.A.G.~Monard$^{28}$,
E.~Ofek$^{29}$,
R.~W.~Pogge$^{4}$, \\ (The $\mu$FUN Collaboration)\\
R.A.~Street$^{8}$,
D.M.~Bramich$^{30}$,
M.~Dominik$^{31}$,
K.~Horne$^{31}$,
C.~Snodgrass$^{32,33}$,
I.A.~Steele$^{34}$,  \\ (The Robonet Collaboration)\\
M.D.~Albrow$^{35}$,
E.~Bachelet$^{8}$,
V.~Batista$^{9}$,
J.-P.~Beaulieu$^{9}$,
S.~Brillant$^{36}$,
J.A.R.~Caldwell$^{37}$,
A.~Cole$^{11}$,
C.~Coutures$^{9}$,
S.~Dieters$^{11}$,
D.~Dominis~Prester$^{38}$,
J.~Donatowicz$^{39}$,
P.~Fouqu\'e$^{40,41}$,
M.~Hundertmark$^{31,42}$,
U.~G.~J{\o}rgensen$^{42}$,
N.~Kains$^{15}$,
S.R.~Kane$^{43}$,
J.-B.~Marquette$^{9}$,
J.~Menzies$^{44}$,
K.R.~Pollard$^{35}$,
C.~Ranc$^{8}$,
K.C.~Sahu$^{15}$,
J.~Wambsganss$^{45}$,
A.~Williams$^{46,47}$, and
M.~Zub$^{45}$
\\ (The PLANET Collaboration)
 } 
              

\affil{$^{1}$Code 667, NASA Goddard Space Flight Center, Greenbelt, MD 20771, USA;    \\ Email: {\tt david.bennett@nasa.gov}}
\affil{$^{2}$Deptartment of Physics,
    University of Notre Dame, 225 Nieuwland Science Hall, Notre Dame, IN 46556, USA; }
\affil{$^{3}$Warsaw University Observatory, Al.~Ujazdowskie~4, 00-478~Warszawa,Poland}
\affil{$^{4}$Dept.\ of Astronomy, Ohio State University, 140 West 18th Avenue, Columbus, OH 43210, USA}
\affil{$^{5}$Max-Planck-Institute for Astronomy, K\"onigstuhl 17, 69117 Heidelberg, Germany}
\affil{$^{6}$Korea Astronomy and Space Science Institute, Daejon 34055, Republic of Korea}
\affil{$^{7}$Astronomisches Rechen-Institut, Zentrum f{\"u}r Astronomie der Universit{\"a}t Heidelberg (ZAH), 69120 Heidelberg, Germany}
\affil{$^{8}$Las Cumbres Observatory Global Telescope Network, 6740 Cortona Drive, suite 102, Goleta, CA 93117, USA}
\affil{$^{9}$Institut d'Astrophysique de Paris, 98 bis bd Arago, 75014 Paris, France}
\affil{$^{10}$Institute of Natural and Mathematical Sciences, Massey University, Auckland 0745, New Zealand}
\affil{$^{11}$School of Math and Physics, University of Tasmania, Private Bag 37, GPO Hobart, 7001 Tasmania, Australia}
\affil{$^{12}$Department of Physics, University of Auckland, Private Bag 92019, Auckland, New Zealand}
\affil{$^{13}$Department of Astronomy, Yale University, New Haven, CT 06511, USA}
\affil{$^{14}$Pennsylvania State University, 537 Davey Lab, University Park, PA 16802, USA}
\affil{$^{15}$Space Telescope Science Institute, 3700 San Martin Drive, Baltimore, MD 21218, USA}
\affil{$^{16}$Institute for Space-Earth Environmental Research, Nagoya University, Nagoya 464-8601, Japan}
\affil{$^{17}$Okayama Astrophysical Observatory, National Astronomical Observatory of Japan, 3037-5 Honjo, Kamogata, Asakuchi, Okayama 719-0232, Japan}
\affil{$^{18}$Department of Earth and Space Science, Graduate School of Science, Osaka University, Toyonaka, Osaka 560-0043, Japan}
\affil{$^{19}$Nagano National College of Technology, Nagano 381-8550, Japan}
\affil{$^{20}$Tokyo Metropolitan College of Aeronautics, Tokyo 116-8523, Japan}
\affil{$^{21}$School of Chemical and Physical Sciences, Victoria University, Wellington, New Zealand}
\affil{$^{22}$Mt.\ John University Observatory, P.O. Box 56, Lake Tekapo 8770, New Zealand}
\affil{$^{23}$Department of Physics, Faculty of Science, Kyoto Sangyo University, 603-8555 Kyoto, Japan}
\affil{$^{24}$Vintage Lane Observatory, Blenheim, New Zealand}
\affil{$^{25}$Department of Physics, Texas A\&M University, 4242 TAMU, College Station, TX 77843-4242, USA}
\affil{$^{26}$Department of Particle Physics and Astrophysics, Weizmann Institute of Science, 234 Herzl St.\ 76100 Rehovot Israel}
\affil{$^{27}$Department of Physics, Chungbuk National University, Cheongju 361-763, Republic of Korea}
\affil{$^{28}$Bronberg and Kleinkaroo Observatories, Centre for Backyard Astrophysics, Calitzdorp, South Africa}
\affil{$^{29}$Weizmann Institute of Science, 234 Herzl Street, Rehovot 7610001 Israel}
\affil{$^{30}$Qatar Environment and Energy Research Institute(QEERI), HBKU, Qatar Foundation, Doha, Qatar}
\affil{$^{31}$SUPA, School of Physics \& Astronomy, University of St Andrews, North Haugh, St Andrews KY16 9SS, UK}
\affil{$^{32}$Planetary and Space Sciences, Department of Physical Sciences, The Open University, Milton Keynes, MK7 6AA, UK}
\affil{$^{32}$Max Planck Institute for Solar System Research,Justus-von-Liebig-Weg 3, 37077 G{\"o}ttingen, Germany }
\affil{$^{34}$Astrophysics Research Institute, Liverpool John Moores University, Liverpool CH41 1LD, UK}
\affil{$^{35}$University of Canterbury, Dept. of Physics and Astronomy, Private Bag 4800, 8020 Christchurch, New Zealand}
\affil{$^{36}$ESO Vitacura, Alonso de C—rdova 3107. Vitacura, Casilla 19001, Santiago 19, Chile}
\affil{$^{37}$McDonald Observatory, 82 Mt Locke Rd, McDonald Obs TX 79734 USA}
\affil{$^{38}$Department of Physics, University of Rijeka, Radmile Matej v{c}i\'{c} 2, 51000 Rijeka, Croatia}
\affil{$^{39}$Technical 2niversity of Vienna, Department of Computing, Wiedner Hauptstrasse 10, 1040 Wien, Austria}
\affil{$^{40}$CFHT Corporation, 65-1238 Mamalahoa Hwy, Kamuela, Hawaii 96743, USA}
\affil{$^{41}$IRAP, CNRS - Universit\'e de Toulouse, 14 av. E. Belin, F-31400 Toulouse, France}
\affil{$^{42}$Niels Bohr Institutet, K{\o}benhavns Universitet, Juliane Maries Vej 30, 2100 K{\o}benhavn {\O}, Denmark}
\affil{$^{43}$Department of Physics and Astronomy, San Francisco State University, 1600 Holloway Avenue, San Francisco, CA 94132, USA}
\affil{$^{44}$South African Astronomical Observatory, PO Box 9, Observatory 7935, South Africa}
\affil{$^{45}$Astronomisches Rechen-Institut, Zentrum f{\"u}r Astronomie der Universit{\"a}t Heidelberg (ZAH), M\"onchhofstra{\ss}e 12-14, 69120 Heidelberg, Germany}
\affil{$^{46}$Perth Observatory, Walnut Road, Bickley, Perth 6076, Australia}
\affil{$^{47}$International Centre for Radio Astronomy Research, Curtin University, Bentley, WA 6102, Australia}
\affil{$^{\dagger}$deceased}
\affil{$^{M}$MOA Collaboration}
\affil{$^{P}$PLANET Collaboration}
\affil{$^{O}$OGLE Collaboration}
\affil{$^{\mu}\mu$FUN Collaboration}
\affil{$^{R}$Robonet Collaboration}


\begin{abstract}
We present the analysis of the first circumbinary planet microlensing event, OGLE-2007-BLG-349.
This event has a strong planetary signal that is best fit with a mass ratio 
of $q \approx 3.4\times10^{-4}$, but there is an additional signal due to an additional
lens mass, either another planet or another star. We find acceptable light
curve fits with two classes of models: 2-planet models (with a single host star) and circumbinary
planet models. The light curve also reveals a significant microlensing
parallax effect, which constrains the mass of the lens system to be $M_L \approx 0.7 \msun$.
Hubble Space Telescope images resolve the lens and
source stars from their neighbors and indicate 
excess flux due to the star(s) in the lens system. This is consistent with the predicted flux from the
circumbinary models, where the lens mass is shared between two stars, but there
is not enough flux to be consistent with the 2-planet, 1-star models. 
So, only the circumbinary models are consistent with the 
HST data. They indicate a planet of mass $m_c = 80\pm 13\,\mearth$, orbiting
a pair of M-dwarfs with masses of $M_A = 0.41\pm 0.07\msun$ and $M_B = 0.30\pm 0.07\msun$,
which makes this the lowest mass circumbinary planet system known. 
The ratio of the separation between the planet and the center-of-mass to the separations
of the two stars is $\sim 40$, so unlike most of the circumbinary planets
found by Kepler, the planet does not orbit near the stability limit.
\end{abstract}


\section{Introduction}
\label{sec-intro}

One of the main features of the observational study of extrasolar planets has been
the continuing stream of surprise observational discoveries. These include planets orbiting
a pulsar \citep{pulsar_planets}, hot Jupiters \citep{51peg}, systems of short period,
low-density planets in tightly packed orbits \citep{lissauer_kep11}, and circumbinary
planets \citep{doyle11} close to the stability limit. Circumbinary planets and planets
in close binary systems are very difficult to detect with the radial velocity method, but
Kepler has proved quite adept at finding such systems
\citep{doyle11,welch12,welch15,orosz12,kostov13,kostov14,kostov16}. Gravitational 
microlensing \citep{bennett_rev,gaudi_araa} has demonstrated the ability to detect such systems 
\citep{mps-97blg41,gould14,poleski14,udalski15} (either circumbinary planets or planets
orbiting one member of a relatively close binary). Two of these claimed microlensing
planets in binary systems have turned out to be incorrect,
MACHO-97-BLG-41 \citep{mps-97blg41,albrow-97blg41,jung13} and
OGLE-2013-BLG-0723 \citep{udalski15,han_ob130723},
but this is largely an issue that can be addressed by greater care in event modeling. 
These events still help to establish the sensitivity of the microlensing method
to planets in close binary systems, because in each case, the light curve measurements
do definitively distinguish between the triple-lens, planetary models, and the close
binary models without a planet.

In this paper, we present the first circumbinary planet found by microlensing,
OGLE-2007-BLG-349L(AB)c\footnote{Our designation for this event corrects an apparent 
inconsistency in the naming of planets in binary systems by using a unique letter for
each mass in the system, following the convention for planets orbiting single stars.}. 
The signal for this event is dominated by the microlensing
effect of a Saturn mass ratio planet, but the very central part of the planetary binary
lens light curve does not fit the data. As we show in Section~\ref{sec-lc}, the
light curve can be fit by models with an additional lens mass, either another planet
or another star. However, the light curve data does not tell us which of these models is correct.
Nevertheless, the light curve does reveal finite source effects and a microlensing
parallax signal that allow us to determine the lens system mass, as we discuss in
Section~\ref{sec-lensprop}. 

In Section~\ref{sec-HST}, we present Hubble Space Telescope (HST) observations of the
OGLE-2007-BLG-349 lens system and source star. These observations clearly indicate 
excess flux at the position of the source, which is consistent with the circumbinary models
but not the two-planet models. If the stellar mass of the lens system is divided into
two masses, then it is substantially fainter ($\sim 1.6\,$mag) in the $I$-band
than a single host star would be. And it is only such a faint lens system that is consistent
with the HST images, and so it is the HST observations that select the circumbinary
model over the two-planet models. In Section~\ref{sec-HSTmod}, we add the
lens brightness constraint to our light curve modeling in order to confirm this
conclusion, and we find that two-planet models with an extremely faint host star
(presumably a white dwarf) do better than the best two-planet models with a 
main sequence host star. But, these models are still substantially worse than
the circumbinary models, so they are excluded.

We consider adaptive optics observations of the source and lens stars in 
Section~\ref{sec-NACO}, and we find that these observations provide marginal
support for the circumbinary interpretation of this light curve. Finally, in 
Section~\ref{sec-conclude}, we discuss the implications of this discovery
for our understanding of the properties of exoplanets.

\section{Light Curve Data and Photometry}
\label{sec-lc_data}

Microlensing event OGLE-2007-BLG-349, at ${\rm RA} =18$:05:24.43, 
${\rm  DEC} = -26$:25:19.0, and Galactic coordinates $(l, b) = (4.3802, -2.5161)$, was 
identified as a microlensing candidate by the Optical Gravitational Lensing
Experiment (OGLE) Collaboration Early Warning System (EWS) 
\citep{ogle-ews} and announced on 2007 July 2. Later that month, the event was
independently identified and announced by the MOA Collaboration as MOA-2007-BLG-379.
In mid-to-late August, this event was 
recognized as a potential high magnification event, with high sensitivity to planets, by the
$\mu$FUN, Robonet and the PLANET microlensing follow-up groups, so they started observations
prior to peak magnification. On 2007 Sep 4, the planetary anomaly was first identified in the 
OGLE data by the $\mu$FUN and OGLE groups at 
${\rm HJD}^\prime = {\rm HJD} - 2450000 = 4348.5$.
Despite the fact that this event occurred near the end of the Galactic bulge observing
season, we achieved nearly complete 
coverage of the light curve peak, with the largest data gap of only 55 minutes over
a period of 22 hours centered on the light curve peaks. This coverage was achieved with
the combined data of the OGLE and MOA survey groups and the $\mu$FUN,
Robonet, and PLANET microlensing follow-up groups.

The data set we use in this analysis consists of microlensing survey data from the
OGLE 1.3m telescope in Chile in the $I$-band and the MOA 1.8m telescope in 
New Zealand in the custom MOA-$R$-band, which is equivalent to the sum of
the Cousins $R$+$I$-bands, as well as data from 6 telescopes operated by
microlensing follow-up groups. Four of these telescopes are operated by
the Microlensing Follow-up Network ($\mu$FUN). $\mu$FUN provided 
$V$, $I$, and $H$-band data from the 1.3m SMARTS telescope at CTIO in Chile, $I$-band
data from the 1.5m Palomar telescope in California, and 
unfiltered data from the 0.35m Bronberg Observatory telescope in South Africa
and the 0.4m Vintage Lane Observatory (VLO) telescope in New Zealand. The 
RoboNet Collaboration has provided $R$-band data from the 2m Faulkes North
Telescope (FTN), and the Probing Lensing Anomalies NETwork (PLANET)
Collaboration provided $I$-band data from the 1.0m Canopus Observatory telescope.
We exclude from the analysis data from several $\mu$FUN 
observatories that were unable to obtain data near the light curve peak.

The data were reduced with various implementations of the difference imaging
method \citep{tom96}. The MOA and OGLE data were reduced with their
respective pipelines \citep{bond01,ogle-pipeline}. The PLANET data were reduced
with a version of ISIS \citep{ala98}, and the RoboNet data were reduced with the
RoboNet pipeline \citep{bramich08}. Most of the $\mu$FUN data were reduced with
the OGLE pipeline, but the CTIO $H$-band data were reduced with PySIS 
\citep{albrow-pysis}.

\begin{deluxetable}{ccc}
\tablecaption{Error Bar Modification Parameters\label{tab-eparam}}
\tablewidth{0pt}
\tablehead{
\colhead{Data Set}  & \colhead{$K$} & \colhead{$\sigma_{\rm min}$}  }
\startdata 
OGLE-$I$ & 0.979 & 0.006 \\
MOA-($R$+$I$) & 0.932 & 0.007 \\
CTIO-$I$ & 1.500 & 0.003 \\
CTIO-$H$ & 2.142 & 0.003 \\
FTN-$R$ & 2.598 & 0.003 \\
Palomar-$I$ & 1.605 & 0.004  \\
Canopus-$I$ & 0.974 & 0.002  \\
Bronberg-$Un$ & 1.048 & 0.012 \\
VLO-$Un$ & 1.312 & 0.008  \\
\enddata
\tablecomments{ Passband $Un$ refers to unfiltered imaging. }
\end{deluxetable}

We follow the usual method \citep{yee12} to improve
the photometric error bars with the following formula:
\begin{equation}
\label{eq-errorbar}
\sigma = K\sqrt{\sigma_0^2+ \sigma_{\rm min}^2}\ ,
\end{equation}
where $\sigma_0$ is the error bar estimate provided by the photometry code. 
The error bars in equation~\ref{eq-errorbar} are in linear units as a fraction of
the measured flux. For photometry provided in magnitudes, the error bars are
converted to linear units prior to the equation~\ref{eq-errorbar} modifications.
The error bar correction parameters $K$ and $\sigma_{\rm min}$ for each data set are 
listed in Table~\ref{tab-eparam}. These error bar modifications are made based on
an approximately correct reference model to give $\chi^2/{\rm d.o.f} = 1$ for
each data set. The corrected error bars are normally then used to make more
accurate estimates for the uncertainties in the physical parameters of
the lens system, and the selection of the correct model does not depend on
the error bar corrections. In this case, however, there are competing models,
so one might be concerned that the final conclusions could be dependent on
which model light curve is used to determine the error bar modification
parameters. Fortunately, in this case, the competing two-planet and circumbinary
light curves are so similar that the choice of the reference model 
does not have a significant effect on our analysis. The error bar modifications
are essentially independent of the reference model.

\section{Light Curve Models}
\label{sec-lc}

\begin{figure}
\epsscale{0.7}
\plotone{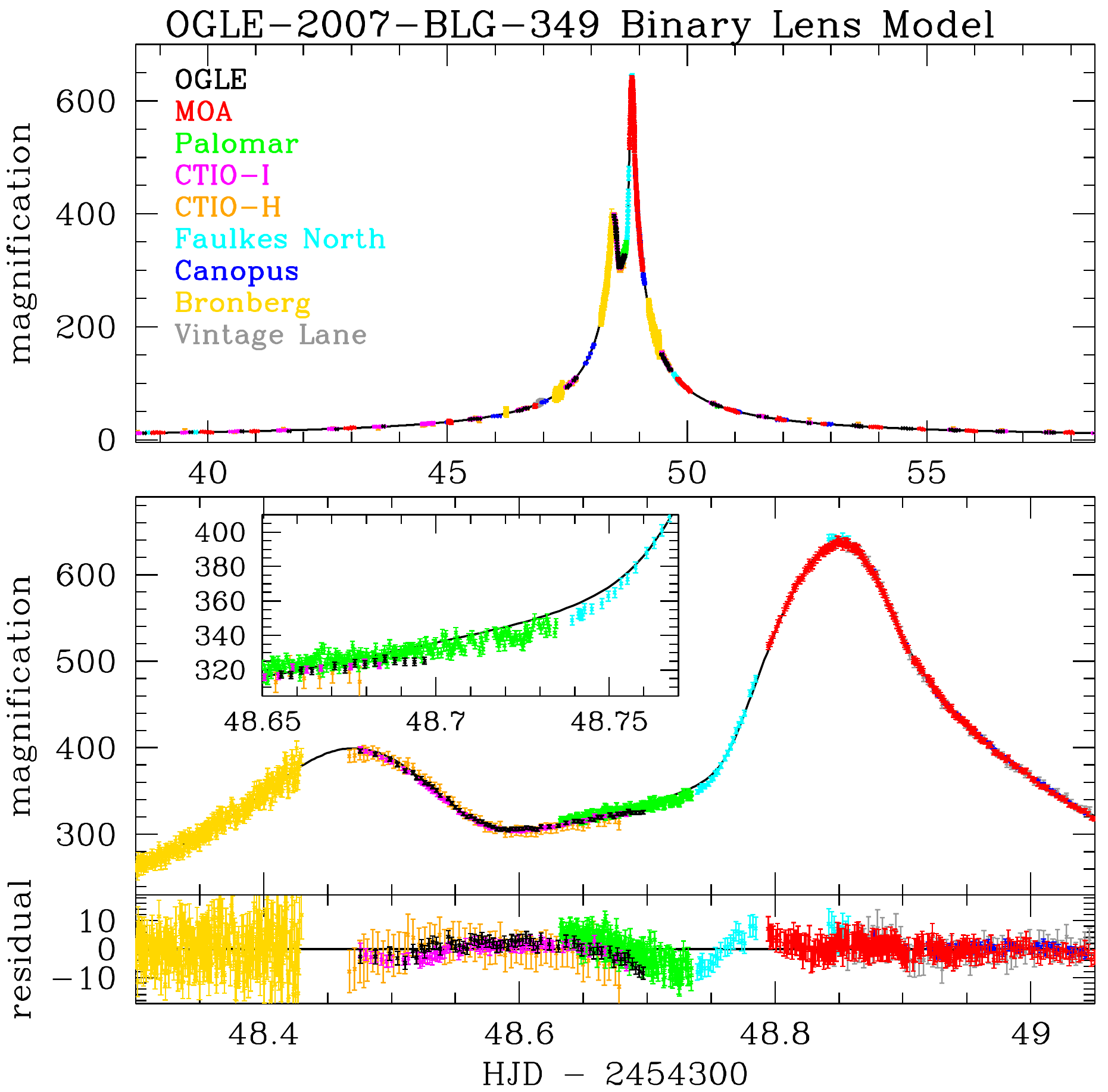}
\caption{Best fit 1-star plus 1-planet light curve. Top panel shows the 20 days
centered on the light curve peak, and the middle and bottom panels show
the light curves and the residuals for the central day of the light curve.
The single planet model matches all the major light curve features, but there are 
significant residuals at $ 48.65 \simlt {\rm HJD}-2454300 \simlt 48.82$. This is
quite close to the $t_0$ value of the fit, which is the time when a light curve
feature due to an additional mass would be expected.
\label{fig-lc_1pl}}
\end{figure}

The preliminary modeling of this event was done independently using the methods of
\citet{dong06,dong-moa400}
and \citet{bennett-himag} to first search for the parameters of the planet that
dominates the anomaly signal.
This light curve is strongly dominated by the signal of a Saturn mass-ratio planet
with parameters quite similar to the model circulated by one of us (DPB) within
24 hours of the first detection of the planetary anomaly. (The basic geometry of
the event was identified even earlier by two of us: AC and NJR.) 

\begin{deluxetable}{ccccc}
\tablecaption{Best Fit Model Parameters
                         \label{tab-bestmparam} }
\tablewidth{0pt}
\tablehead{
\colhead{parameter}  & \colhead{units} &
\colhead{1-planet} & \colhead{2-planet}  & \colhead{circumbinary} 
}  

\startdata
$t_E$ & days & 116.703 & 113.520 & 117.720 \\
$t_0$ & ${\rm HJD}^\prime$ & 4348.7534 & 4348.7469 & 4348.7465 \\
$u_0$ & & -0.0021516 & 0.0020581 & -0.0019818 \\
$d_{1\rm cm}$ & & 1.25268 & 0.79607 & 0.81468 \\
$d_{23}$ & & - & 0.95046 & 0.019905 \\
$\theta_{1\rm cm}$ & radians & 4.40140 & 1.89437 & 4.35940 \\
$\phi_{23}$ & radians & - & -3.07611 & 0.36989 \\
$\epsilon_1$ & $10^{-4}$ & 3.7794 & 3.7669 & 3.4099 \\
$\epsilon_2$ & & 0.999622 & $8.5025\times 10^{-6}$ & 0.46479 \\
$\epsilon_3$ & & - & 0.999615 & 0.53487 \\
$t_\ast$ & days & 0.06614 & 0.06930 & 0.07064 \\
$\dot{d}_{23x}$ & ${\rm days}^{-1}$ & - & 0.0 & 0.010478 \\
$\dot{d}_{23y}$ & ${\rm days}^{-1}$ & - & 0.0 & -0.006360 \\
$1/T_{\rm orb}$ & ${\rm days}^{-1}$ & - & 0.0 & 0.059380 \\
$\pi_E$ & & 0.09693 & 0.19070 & 0.17458 \\
$\phi_E$ & radians & 1.69255 & 2.50170 & 0.62638 \\
$\theta_E$ & mas & 1.1828 & 1.1158 & 1.1138 \\
$M_L$ & $\msun$ & 1.4984 & 0.7185 & 0.7835 \\
$I_{S}$ &  mag & 20.357 & 20.324 & 20.365 \\
$I_L$ &  mag & - & 19.634 & 21.465 \\
$I_{SL}(t_{H2})$ & mag & - & 19.162 & 20.009 \\
$V_S$ & mag  & 22.367 & 22.334 & 22.375 \\
$H_S$ & mag  & 18.260 & 18.226 & 18.267 \\
fit $\chi^2$ & & 4237.56 & 3382.64 & 3382.25 \\
dof & & 3571 & 3568 & 3566 \\
\enddata
\tablecomments{${\rm HJD}^\prime = {\rm HJD}-2,450,000$. The reference time for the
microlensing parallax and orbital motion parameters is $t_{\rm fix} = 4349$.}

\end{deluxetable}

The best fit binary lens (1 star + 1 planet) model is shown in Figure~\ref{fig-lc_1pl},
and the parameters of this model are given in Table~\ref{tab-bestmparam}, including
a significant microlensing parallax signal. (We use polar coordinates for the
microlensing parallax vector, such that $\pi_{E,N} = \pi_E\cos \phi_E$ and
$\pi_{E,E} = \pi_E\sin \phi_E$.)
This model provides a good fit to most of the light curve peak, but it does 
not fit the central part of the light curve at $ 48.65 \simlt {\rm HJD}-2454300 \simlt 48.82$,
or $t\approx t_0$. This is the part of the light curve where we would expect to see the
signal of another lens mass: a second planet or a stellar binary companion to the 
host star. So, we performed another initial condition grid search to explore possible
triple-lens models. The triple lens modeling was made possible by the theoretical
work of \citet{rhie97,rhie02}, which was particularly important for the modeling of
orbital motion in a triple lens system \citep{bennett-ogle109}.
We fixed the parameters describing the best fit planetary binary model
and did the grid search over the parameters that describe the additional mass. There
are three additional parameters for triple lens models:
two parameters describing the position of the third mass and one parameter
describing its mass fraction. Using both methods, we
search for three categories of solutions: 2-planet models (with a single host star),
models with the planet orbiting one member of a wide stellar binary, and models with
a close stellar binary orbited by a circumbinary planet. Our initial triple lens fits were done
with static models, but the period of the stellar binaries for the
circumbinary planet models will only be $\sim 10\,$days. Since the duration of the
light curve peak is $\sim 0.5\,$days, binary orbital motion is likely to be important for these 
circumbinary models. So, after finding the best fit static circumbinary models, we include
orbital motion of the two stars for these models. The circumbinary models include 
three additional parameters: the 2-dimensional velocity in the plane of the sky
and the inverse of the orbital period. But, these orbital parameters are also
subject to a constraint, described below in Section~\ref{sec-lensprop}.

\begin{figure}
\epsscale{1.1}
\plottwo{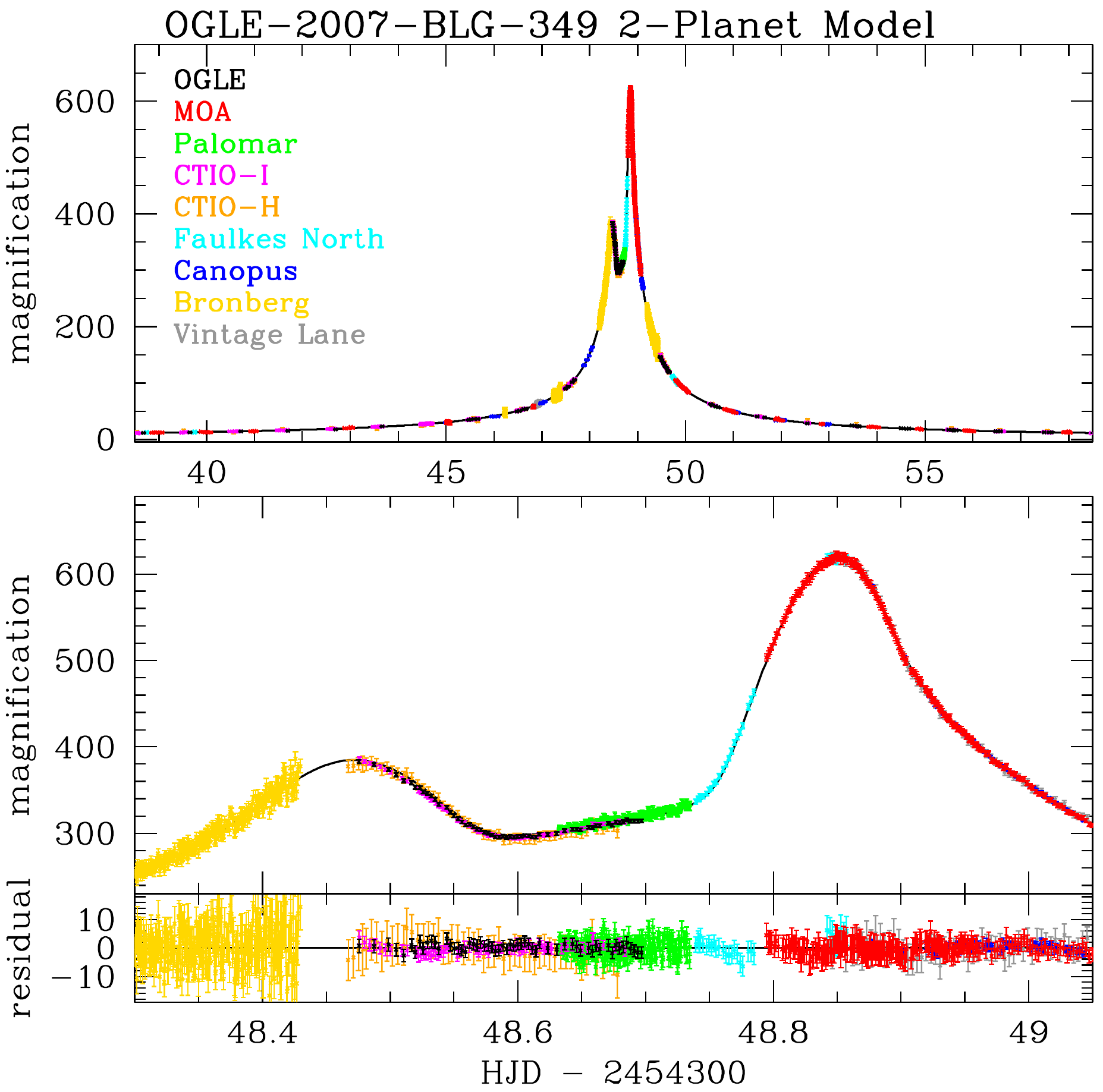}{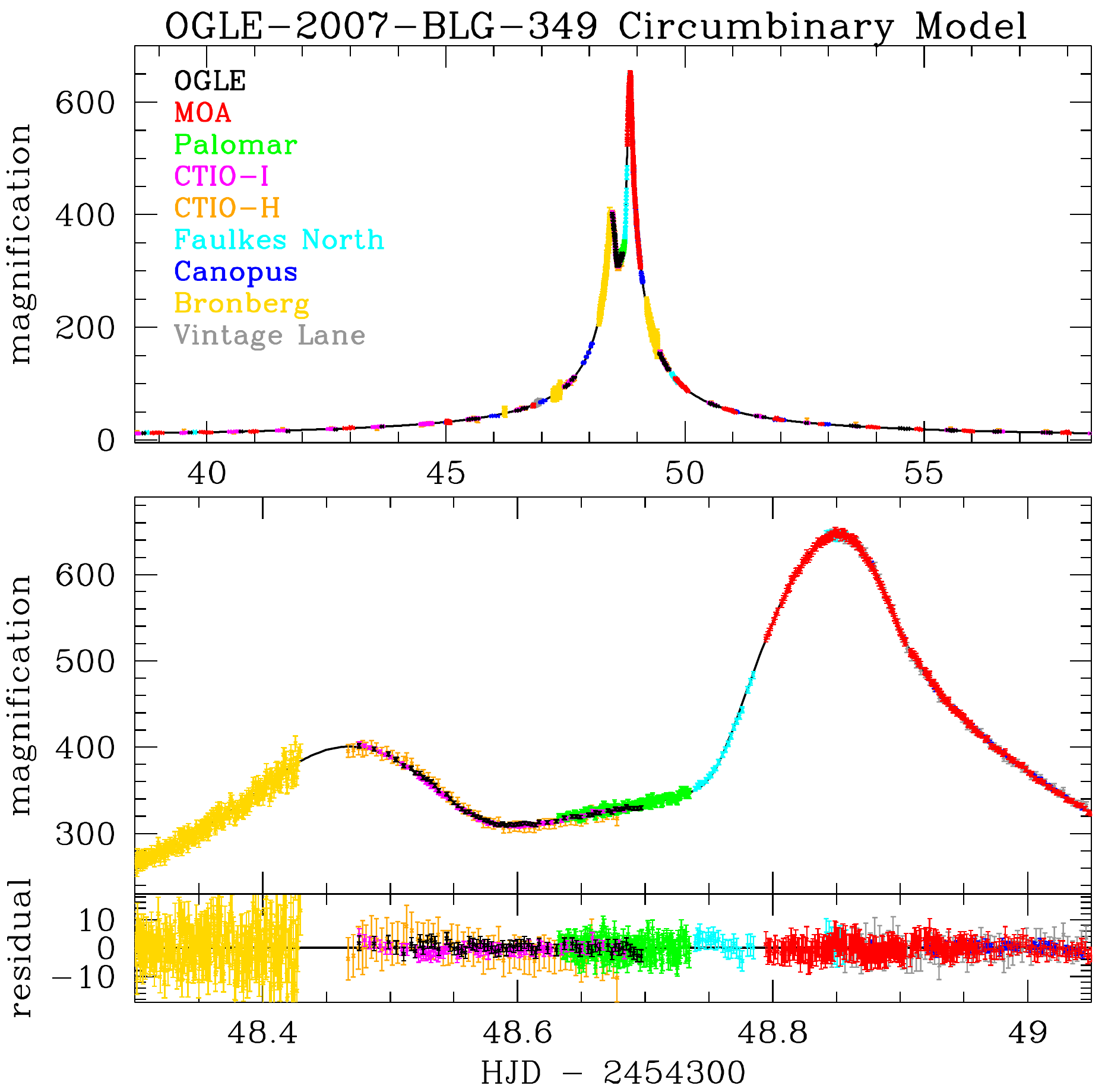}
\caption{Best fit 2-planet (left) and circumbinary (right) light curves with the parameters given in 
Table~\ref{tab-bestmparam}. Both models fit the light curve almost equally well.
\label{fig-lc_2pl_circumbin}}
\end{figure}

All three categories of models that we explore can provide a substantial improvement to 
the light curve over the single planet model, but only the 2-planet and circumbinary models
(with orbital motion) can provide a good fit to the light curve data. The best fit 2-planet
and circumbinary model parameters are given in Table~\ref{tab-bestmparam}, and the
best fit light curves are shown in Figure~\ref{fig-lc_2pl_circumbin}.
The parameters we use are the same as used in the analysis of the 
first triple lens microlensing event,
the two-planet event, OGLE-2006-BLG-109 \citep{gaudi-ogle109,bennett-ogle109}.
The coordinates are based on the center-of-mass system, with a system of total mass $M$.
The length parameters are normalized by the Einstein radius of this total system mass, 
$R_E = \sqrt{(4GM/c^2)D_Sx(1-x)}$, where $x = D_L/D_S$ and $D_L$ and $D_S$ are
the lens and source distances, respectively. ($G$ and $c$ are the gravitational constant
and speed of light, as usual.) $t_E$ is the Einstein radius crossing time, while $t_0$ and
$u_0$ are the time and separation of closest approach of the source to the center-of-mass.
The separation between mass-1 and the center-of-mass of masses 2 and 3 is given by 
$d_{1\rm cm}$, and $d_{23}$ is the distance between masses 2 and 3.
The lens axis is defined as the vector between mass 1 and the mass 2+3 center-of-mass.
and $\theta_{1\rm cm}$ is the angle between the source trajectory and the lens axis,
while $\phi_{23}$ is the angle between the line connecting masses 2 and 3 and the 
lens axis. The mass fractions of each of the 3 masses are $\epsilon_1$, $\epsilon_2$, and
$\epsilon_3$, but these parameters are not independent since 
$\epsilon_1 + \epsilon_2 + \epsilon_3 \equiv 1$. The source radius crossing time is given
by $t_*$. Microlensing parallax is described by $\pi_E$ and $\phi_E$, as described above.
The orbital motion of masses 2 and 3 is described by three parameters.
The instantaneous velocity along the lens axes is given by $\dot{d}_{23x}$, while
$\dot{d}_{23y}$ gives the velocity perpendicular to the lens axis. 
The orbits are constrained to be circular with a period of $T_{\rm orb}$, and 
we use $1/T_{\rm orb}$ as a fit parameter. (In our models, mass-1 refers to the
primary planet, mass-3 refers to a host star and mass-2 can either be a second
host star or a second planet.)

\begin{figure}
\plottwo{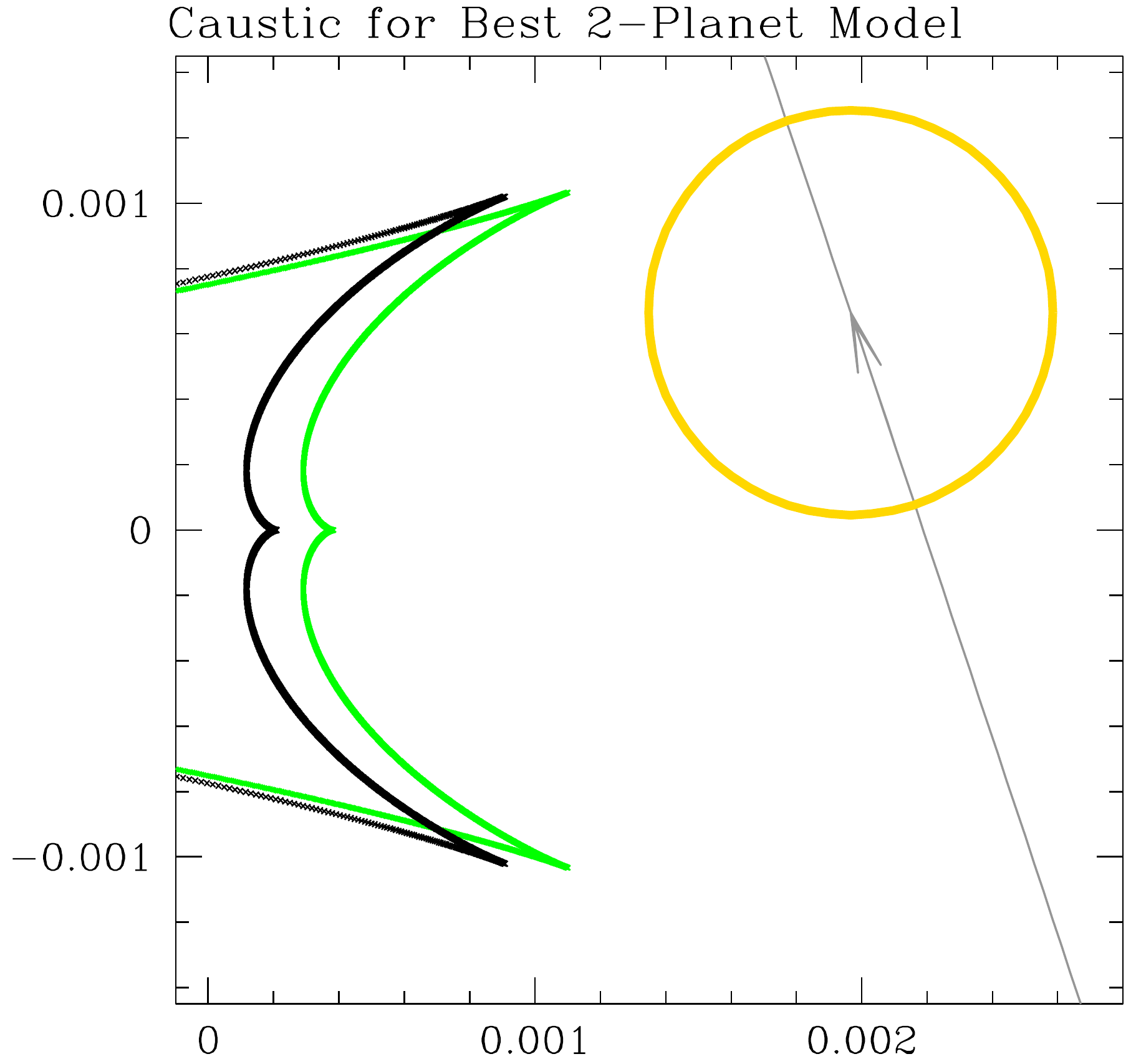}{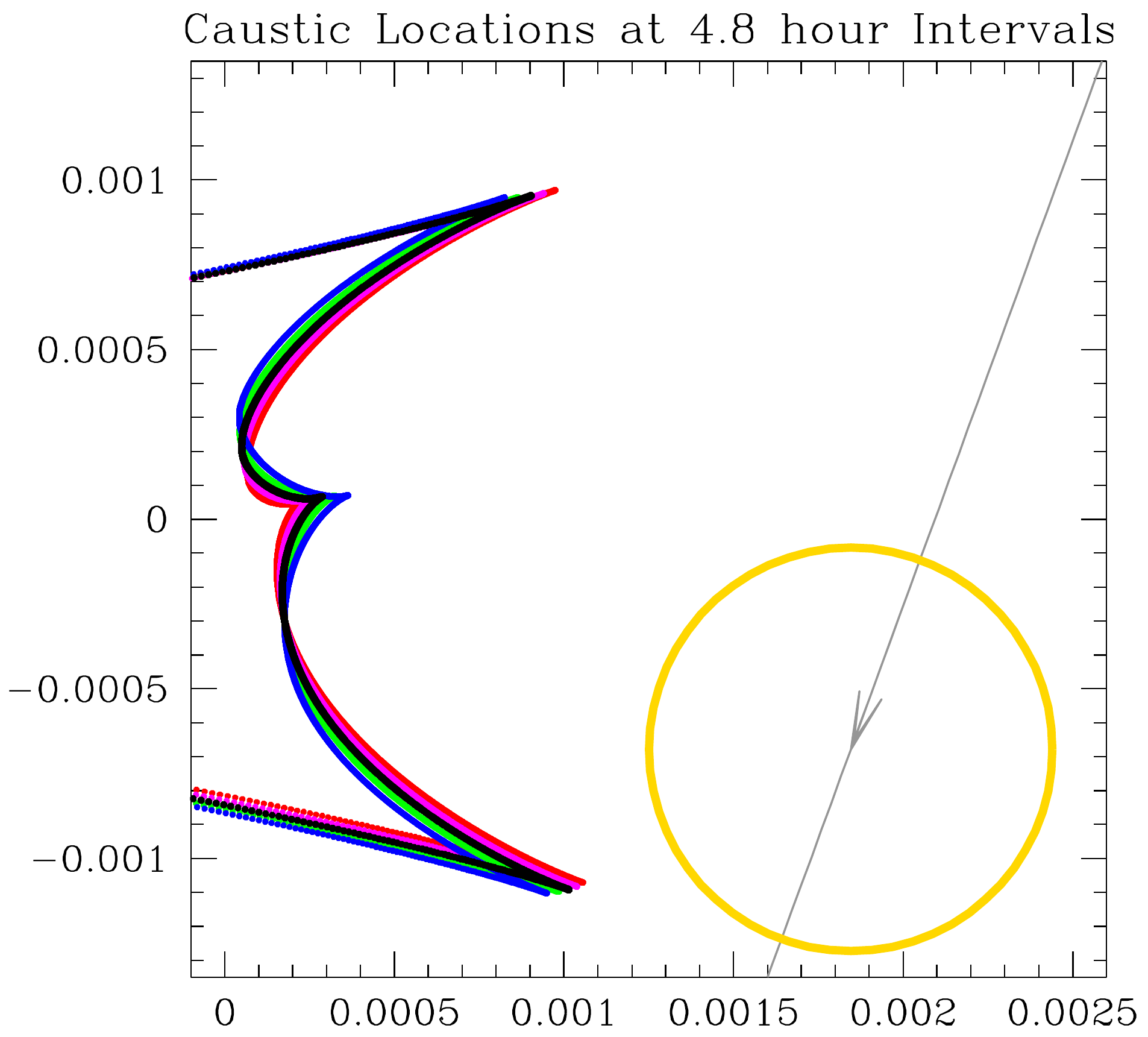}
\caption{OGLE-2007-BLG-349 caustics for the best 1-planet model (in green)
and the best 2-planet model (in black) on the left and for the best fit 
circumbinary planet model at $4.8\,$hr intervals on the right. The gold circle
indicates the source size, and the gray line indicates the source trajectory,
with an arrow indicating the source position at $t = t_0$ and direction of
motion.
\label{fig-caustic}}
\end{figure}

As can be seen from Table~\ref{tab-bestmparam} and Figure~\ref{fig-lc_2pl_circumbin}, 
the light curve data do not distinguish between the 
best 2-planet and circumbinary models. The $\chi^2$ values for the two models are nearly
identical, with the best circumbinary model favored over the best 2-planet model
by $\Delta\chi^2 = 0.39$, but the 2-planet model has two more degrees of
freedom, because of the 3 additional orbital parameters and one constraint
to be explained below in Section~\ref{sec-lensprop}. Either model would 
represent a remarkable discovery. This could be the first circumbinary planet
found by microlensing or the first microlens planet with a mass ratio of $< 10^{-6}$
demonstrating microlensing's sensitivity to Earth-mass planets \citep{bennett96}.

Both the 2-planet and circumbinary models appear to fit the light curve data
equally well, but there are subtle differences that are apparent in the 
residuals plotted in the bottom panel of each light curve figure. These 
residual panels also reveal low-level systematic discrepancies between the
different data sets. 

Figure~\ref{fig-caustic} shows close-ups of the caustic configuration 
for the three best-fit models with the source trajectories given by the 
gray lines. The orbital motion of the two stars causes the caustics to move
for the circumbinary model. They are are displayed at 4.8 hour intervals
starting at $t = 4348.25$ in units of ${\rm HJD}^\prime = {\rm HJD} - 2450000$. 
The sequence of caustic curves is red, magenta, black, cyan, and blue.

\subsection{Microlensing Parallax}
\label{sec-par}

\begin{figure}
\epsscale{0.7}
\plotone{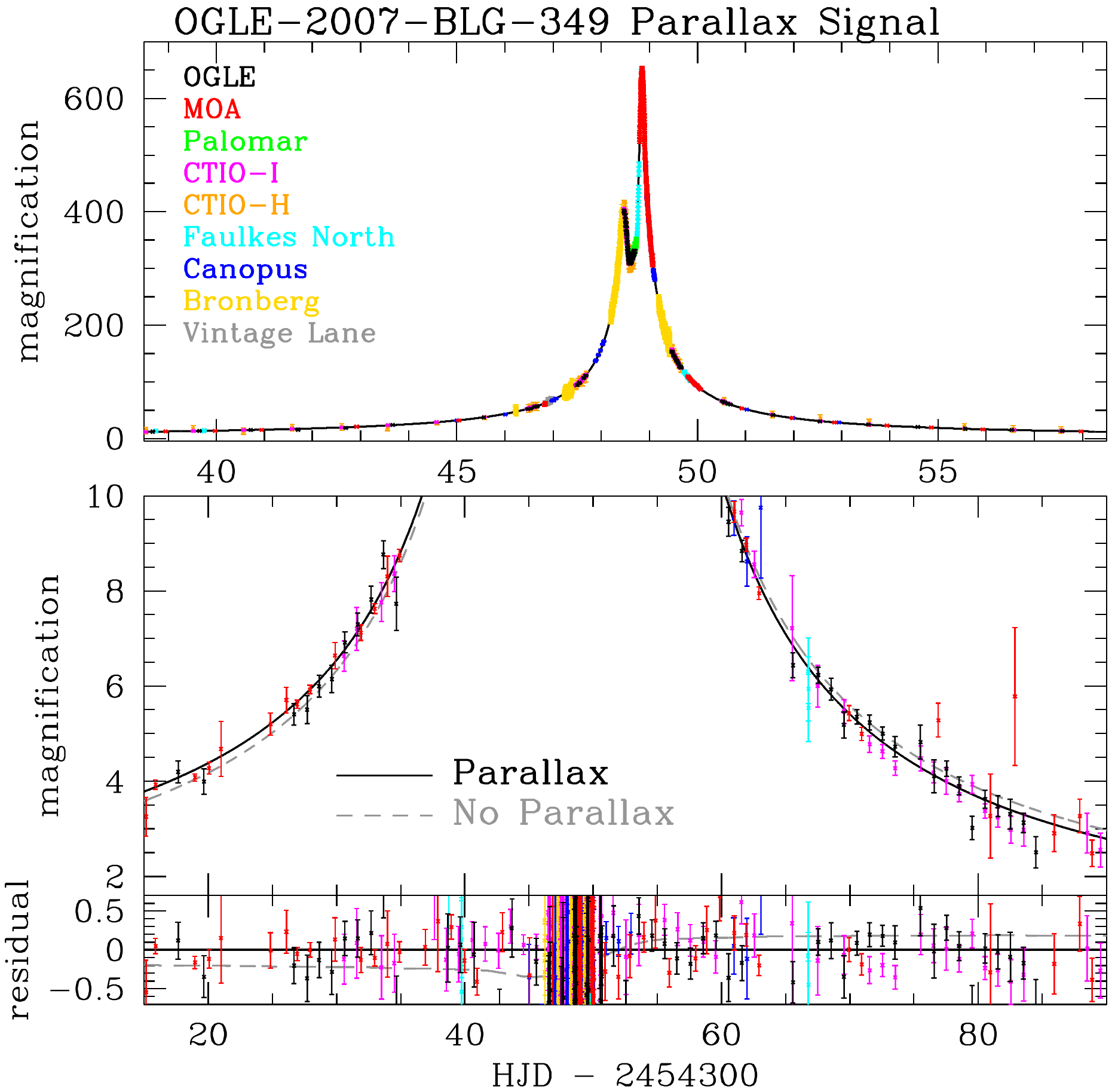}
\caption{Comparison of the best two-planet microlensing model with and without microlensing
parallax plotted as solid black and dashed gray curves, respectively. Much of the parallax 
signal comes in the moderate magnification wings of the light curve. From the bottom panel,
we can see that the data are well above the no-parallax light curve prior to the peak and 
below the no-parallax light curve after the peak.
\label{fig-lc_par2pl}}
\end{figure}

\begin{figure}
\epsscale{0.7}
\plotone{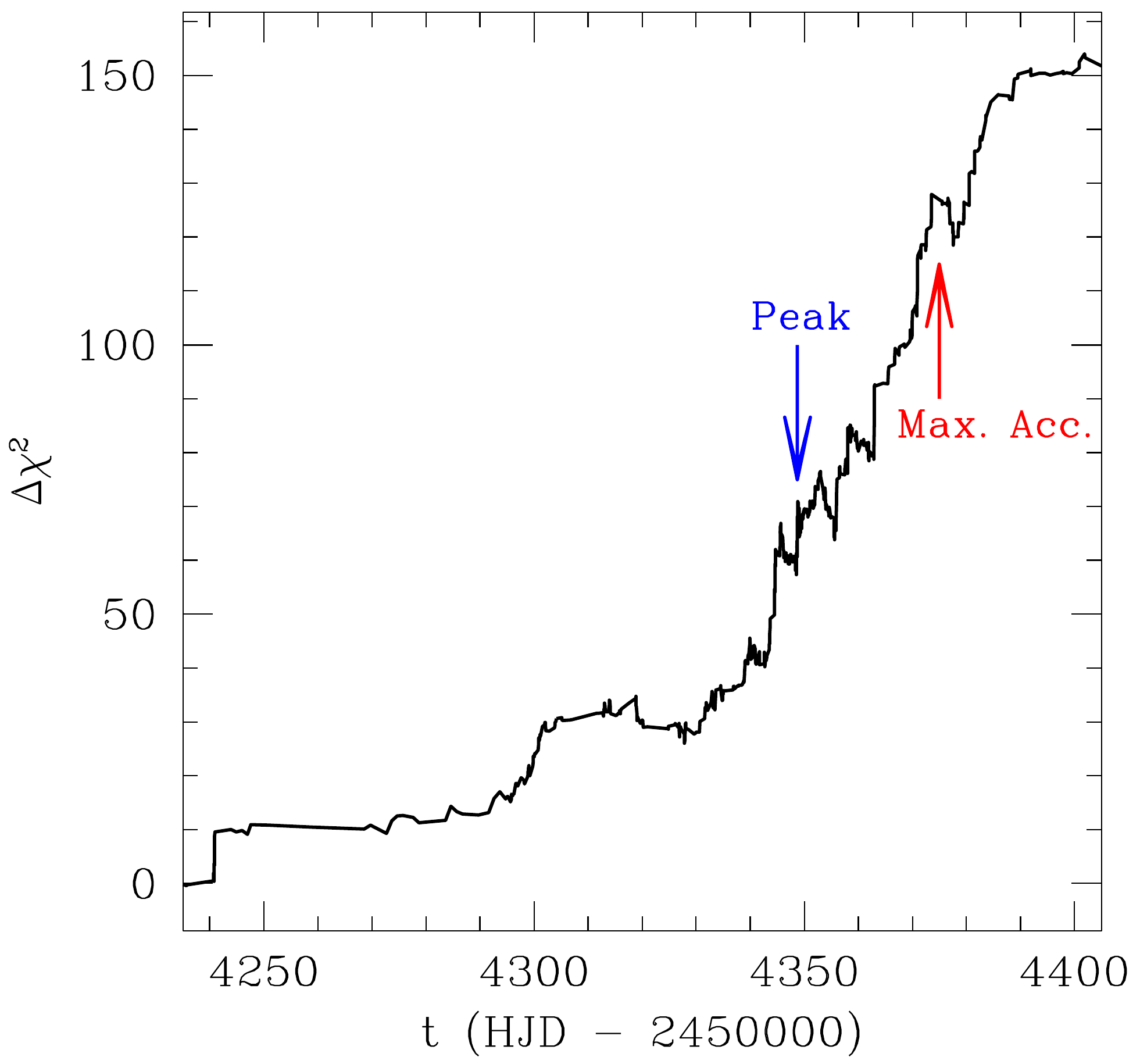}
\caption{The difference in the cumulative $\Delta \chi^2$ between the best fit non-parallax and
parallax circumbinary models as a function of time, with the final $\Delta \chi^2 = 152.8$
for the full light curve. This indicates that the signal is centered between the light curve
peak and the time of maximum acceleration of Earth in the direction perpendicular
to the line-of-sight. This is exactly where we expect the signal to be strongest.
\label{fig-par_dchi2}}
\end{figure}

An important feature of the OGLE-2007-BLG-349 light curve is the microlensing
parallax signal. 
We find that the microlensing parallax effect improves the $ \chi^2 $ by
$\Delta \chi^2 = 152.8$ as indicated in Figures~\ref{fig-lc_par2pl}. 
The parallax signal is quite clear in the second and third panels of this figure.
Figure~\ref{fig-par_dchi2} shows the cumulative difference between the best
fit parallax and non-parallax models. This indicates that the parallax signal
is centered between the time of the peak and the time of the maximum acceleration
of Earth (by the Sun) in the directions perpendicular to the line-of-sight, as expected
for a real microlensing parallax signal.

There are two contributions to the microlensing parallax signal: orbital parallax
due to the Earth's orbital motion around the Sun \citep{gould-par1,macho-par1}
and terrestrial parallax \citep{gould09}, due to observations from telescopes at different
locations on the Earth. The measurement of orbital parallax is fairly common, particularly
for events like OGLE-2007-BLG-349 with durations $t_E > 100\,$days that occur
near the beginning or end of the Galactic bulge observing season, when the acceleration
of Earth was nearly perpendicular to the line-of-sight to the bulge. (OGLE-2007-BLG-349 reached
peak magnification on 2007 September 5, just about 3 weeks before the acceleration
of Earth is perpendicular to the line-of-sight.) The orbital parallax signal is much
stronger than the terrestrial parallax signal, and is dominated by the three data sets 
which observed the event at modest magnification, MOA, OGLE, and $\mu$FUN-CTIO, with 
$\Delta\chi^2$ values of 46.6, 46.0, and 54.1, respectively. Since the acceleration of
Earth is almost entirely in the East-West direction, the East component of the 
orbital parallax solution is much more strongly constrained than the North
component.

Terrestrial parallax is normally quite difficult to measure because the Einstein 
radius projected to the position of the solar system, $\rep$, is usually a few
AU or more, which is a few $\times 10^5$ larger than the separation of 
telescopes on the ground. For ultra-high magnification events with a relatively large
$\pi_E$ value, like OGLE-2007-BLG-224
\citep{gould09}, with a peak magnification of $A_{\rm max} > 2000$, the signal
can become quite strong. For events like OGLE-2007-BLG-349, presented in this
paper, the terrestrial parallax signal is detectable, but relatively weak. However,
terrestrial parallax does not have the strong East-West bias that orbital parallax has
\citep{muraki11}. With data at or near the light curve peak from 
Northern Hemisphere telescopes, like the Faulkes North Telescope (FTS) in Hawaii,
along with Southern Hemisphere telescopes, like the MOA telescope in New Zealand
and the CTIO and OGLE telescopes in Chile, we have some leverage on the North-South
component of terrestrial parallax. So, the terrestrial parallax helps to constrain the 
North component of $\pi_E$, which is weakly constrained by orbital parallax.

High magnification events usually have several degeneracies. There is a degeneracy
between close and wide solutions with  $d_{1\rm cm} \simeq 0.81$ and $d_{1\rm cm} \simeq 1.23$,
respectively. There is also a degeneracy between $u_0 > 0$ and $u_0 < 0$ solutions
that would be exact if there was no microlensing parallax (representing
the two reflections of the lens plane with respect to the projected orbit of Earth).
In this case, the $u_0 > 0$ are excluded by the terrestrial parallax signal. The best 
$u_0 > 0$ and $u_0 < 0$ solutions have nearly identical $\chi^2$ values when terrestrial
parallax is excluded from the modeling, but the $u_0 > 0$ models are disfavored by
$\Delta\chi^2 = 28$ when we include terrestrial parallax. This difference in $\chi^2$
comes from the FTS, CTIO, MOA, and OGLE telescopes.
We will explore these alternative models in more detail after applying the 
Hubble Space Telescope constraints on the lens system brightness.

\section{Lens System Properties}
\label{sec-lensprop}

For events with measurable microlensing parallax signals, it is possible
to determine the lens system mass if the angular Einstein radius, $\theta_E$, can also be
determined \citep{gould-par1,an02},
\begin{equation}
M_L = {\theta_E c^2 {\rm AU}\over 4G \pi_E}
    = {\theta_E \over (8.1439\,{\rm mas})\pi_E} \msun \ .
\label{eq-M}
\end{equation}
Thus, we require the determination of the angular Einstein radius in order to 
determine the lens system mass. Fortunately, the sharp planetary light curve features 
enable a precise measurement of the source radius crossing time, $t_*$.
This provides a determination of the angular Einstein radius, $\theta_E = \theta_* t_E/t_*$, 
if we know the angular radius of the source star, $\theta_*$,
which can be determined from the dereddened source magnitude and color
\citep{kervella_dwarf,boyajian14,adams16} We determine $\theta_*$ in 
Section~\ref{sec-radius}. The lens system distance can also be determined
from $\pi_E$ and $\theta_E$,
\begin{equation}
D_L={1\over \pi_{\rm E}\theta_{\rm E}+\pi_S} \ ,
\label{eq-Dl}
\end{equation}
assuming that the distance to the source, $D_S = 1/\pi_S$ (and its
parallax, $\pi_S$), is known.

\begin{figure}
\epsscale{1.05}
\plottwo{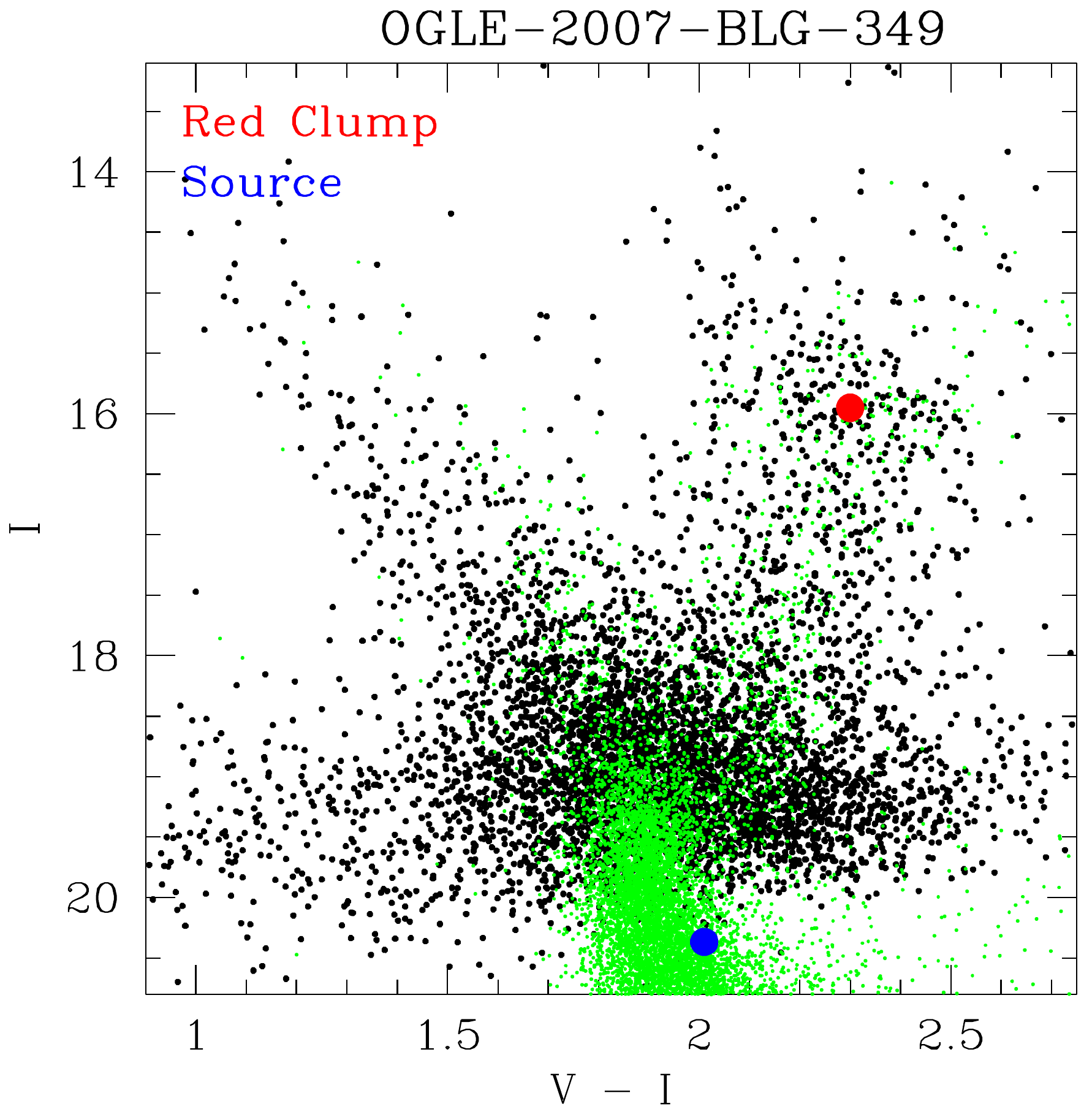}{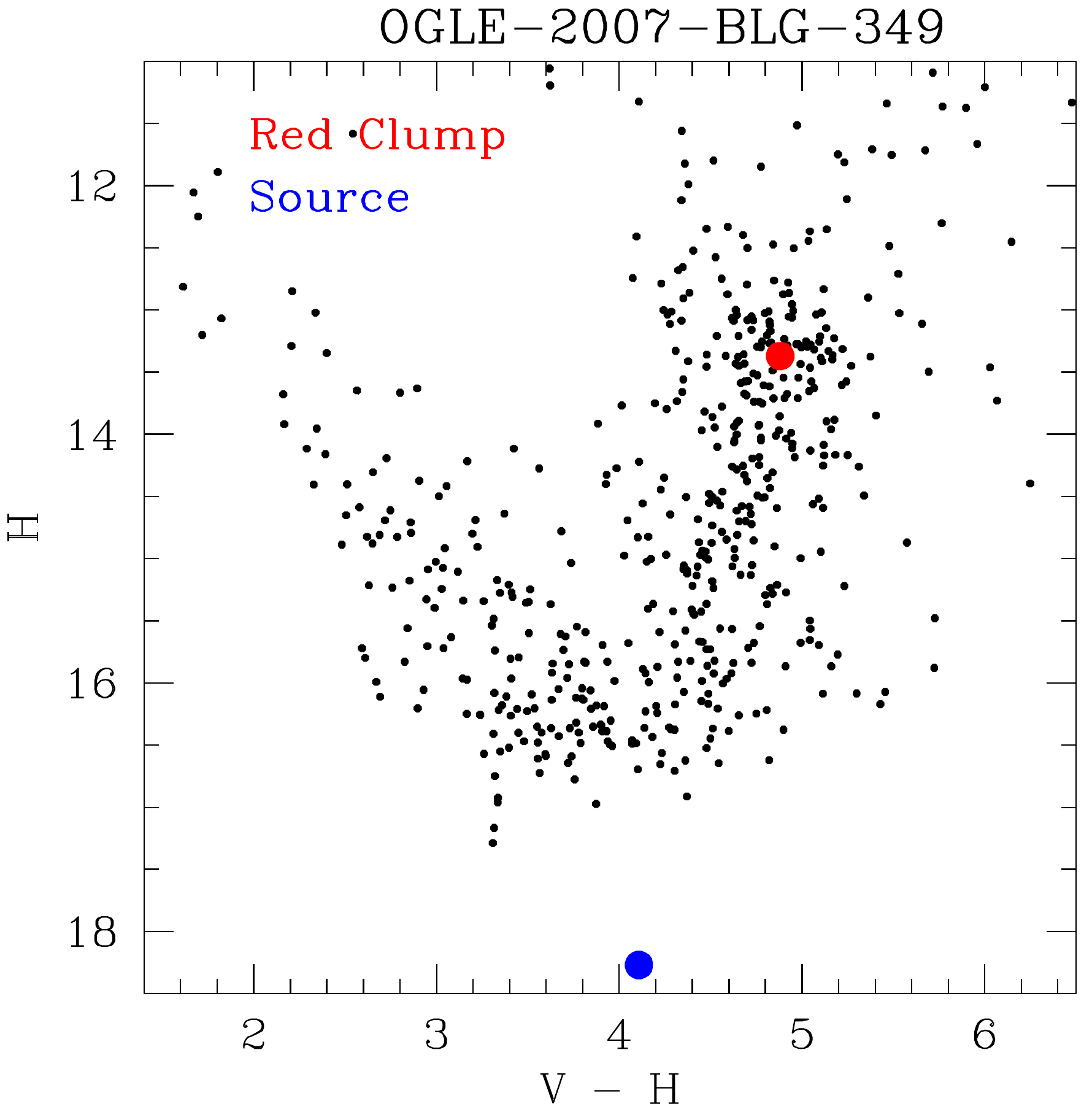}
\caption{The $(V-I,I)$ and $(V-H),H$ color magnitude diagrams (CMD) of the 
stars in the OGLE-III catalog \citep{ogle3-phot} within $90^{\prime\prime}$ of 
OGLE-2007-BLG-349. The green points are the Baade's Window
CMD from \citet{holtzman98} shifted to the same extinction and distance as the 
OGLE-2007-BLG-349 field. The $H$-band magnitudes shown in the right panel 
come from IRSF images that have been calibrated to 2MASS.
The red spots indicate the red clump giant centroid, and the blue spots indicate 
the source magnitudes and colors.
\label{fig-cmd}}
\end{figure}

\subsection{Calibration and Source Radius}
\label{sec-radius}

Figure~\ref{fig-cmd} shows color magnitude diagrams (CMDs) of stars
within $90^{\prime\prime}$ of the OGLE-2007-BLG-349 microlensing
event. The green points in the left panel are from the \citet{holtzman98}
Baade's Window CMD shifted to the same extinction and average
distance as the OGLE-2007-BLG-349 field.
The $V$ and $I$ magnitudes (indicated in black in the left panel)
come from the OGLE-III photometry
catalog \citep{ogle3-phot}, and the $H$-band magnitudes come from
images from the IRSF telescope that have been calibrated to the Two
Micron All Sky Survey (2MASS) 
catalog \citep{2mass_cal}. The stars identified in these IRSF images 
have been cross-matched to the OGLE-III catalog, but not every star
gives a good match. The IRSF images were taken in worse seeing than the OGLE-III
catalog images, so some of the matches between the $VI$ and $H$-band
photometry have uncertainties due to blending where stars resolved in the
OGLE images appear likely to be blended in the IRSF photometry.
We do not include these stars in our $(V-H),H$ CMD, so the number of
stars included in this CMD is smaller than in the $(V-I,I)$ CMD. (The 
OGLE-III $V$-$I$ CMD includes 9421 stars,
but only 317 of the brighter stars have matched one to one with the
stars seen in the $H$-band.)

These CMDs allow us to estimate the extinction toward the field centered
on the source star location. From these CMDs (and the $(I-H),H$ CMD, which is
not shown), we identify the centroid
of the red clump giant distribution at $I_{\rm rc} = 15.95\pm 0.10$,
$(V_{\rm rc}-I_{\rm rc}) = 2.30 \pm 0.05$, $(V_{\rm rc}-H_{\rm rc}) = 4.88 \pm 0.15$
and $(I _{\rm rc}-H_{\rm rc}) = 2.58\pm 0.10$. These are compared 
to the assumed intrinsic (dereddened) properties of red clump giant 
stars \citep{bennett-ogle109,nataf13}, $M_{I\rm rc} = -0.13\pm 0.10$, $(V-I)_{\rm rc0} = 1.06\pm 0.05$, 
$(V-H)_{\rm rc0} = 2.23 \pm 0.07$, and $(I-H)_{\rm rc0} = 1.17 \pm 0.07$. Fitting these constraints
to the \citet{cardelli89} extinction law gives $R_{\rm v} = 3.033$, $A_H = 0.541$, $A_I = 1.818$,
and $A_V = 3.083$.

The $V$ and $I$ source magnitudes were determined by calibrating the CTIO-$V$, and $I$
light curves to the OGLE-III catalog \citep{ogle3-phot} and $H$ source magnitudes were
determined by calibrating to the 2MASS-calibrated IRSF photometry.
The $V$ and $I$ calibrations were done using DoPHOT \citep{dophot}
light curves in order to put them on the same photometric scale as the 
CTIO CMD that was matched to the OGLE-III CMD shown in Figure~\ref{fig-cmd},
while the CTIO $H$-band calibrations were done with a SoDoPHOT reduction
\citep{bennett-sod} for the same reason.
(Note that the OGLE-III light curve photometry is not on the same
scale as the OGLE-III catalog, and an OGLE CMD on the same scale as 
the OGLE-III light curve data was not available.)
The calibrated source magnitudes, $V_S$, $I_S$, and $H_S$, for the best unconstrained
models are displayed in Table~\ref{tab-bestmparam}.

\begin{figure}
\epsscale{0.7}
\plotone{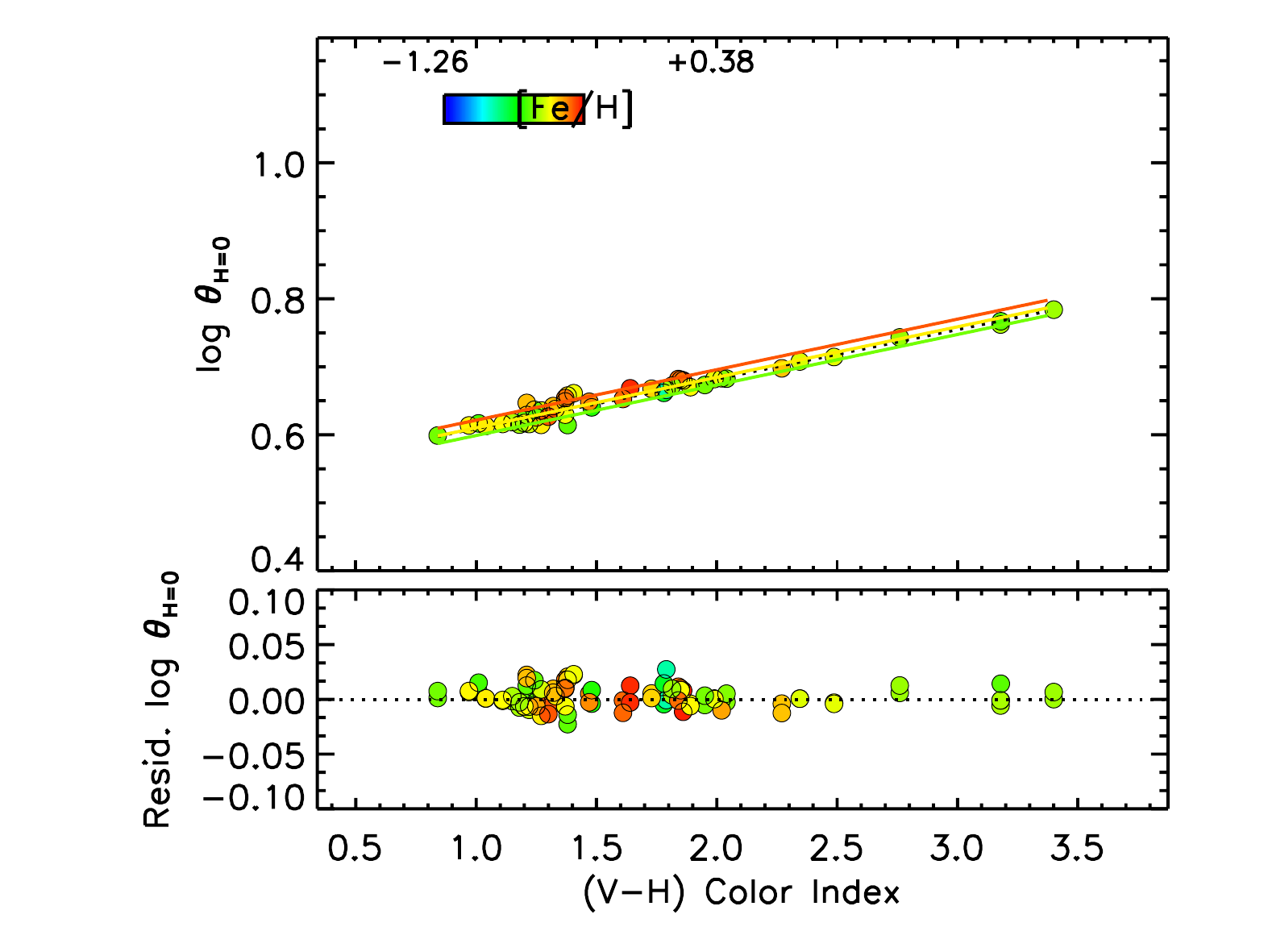}
\caption{The $V-H$, $H$ angular source size relation from the analysis of
\citet{boyajian14}, including the effect of metallicity.
\label{fig-VmH_H_theta}}
\end{figure}

With calibrated source magnitudes and an estimate of the extinction, we are now nearly
ready to determine the angular source radius, using a color-angular-size relation such
as that of \citet{kervella_dwarf} or \citet{boyajian14}, but in fact, we have more information
about the source star. \citet{cohen08} took advantage of the extremely high magnification
of this event to obtain a high resolution spectrum of the source star, when it was magnified
by a factor of $\sim 400$. This allows the metallicity of the source star to be determined, and
we use the determination by \citet{bensby13}, who find $[{\rm Fe/H}] = +0.42 \pm 0.26$. 
This high metallicity is consistent with the CMD location of the source on the red edge of the
bulge main sequence (as represented by source position with respect to the
Baade's Window stars in Figure~\ref{fig-cmd}). Since
metallicity is known to perturb the color-angular-size relations, we asked the authors of
\citet{boyajian14} to derive a relation using the dereddened $H$ and $V$ magnitudes including
the effect of metallicity. The result is shown in Figure~\ref{fig-VmH_H_theta}, which shows the
data and following fit to the data:
\begin{equation}
\log_{10} (2\theta_*/{\rm mas}) = 0.53598 + 0.07427(V_{S0}-H_{S0}) + 0.04511[{\rm Fe/H}] - 0.2H_{S0} \ .
\label{eq-m_thetaE}
\end{equation}
as shown in Figure~\ref{fig-VmH_H_theta}. The subscripts $_{S0}$ indicate extinction-corrected source
magnitudes. If we assume a
1.5\% uncertainty in the model, 2.7\% uncertainty from the [Fe/H] error bar,
$0.1\,$mag uncertainty in $(V_{S0}-H_{S0})$ and $0.02\,$mag calibration uncertainty for $H_{S0}$,
then we find a 3.6\% uncertainty for this relation. (Note that this does not include the
light curve model uncertainty in  $H_{S0}$, which will be handled by a different part of
our analysis.)

Now that we have a formula for the angular source radius, $\theta_*$, we can 
determine that angular Einstein radius, $\theta_E = \theta_* t_E/t_*$ for each light curve
model. This allows us to determine the lens mass, using equation~\ref{eq-M}. The
lens distance can also be determined using Equation~\ref{eq-Dl},
provided that the source distance, $D_S$, is known.

Table~\ref{tab-bestmparam} gives the masses corresponding to the best fit
2-planet and circumbinary models, which are $M_L = 0.7185\msun$ and $0.7835\msun$,
respectively, from equation~\ref{eq-M}.
If we assume a source distance of $8\,$kpc, equation~\ref{eq-Dl} indicates lens
system distances of $D_L = 2.96\,$kpc and $3.13\,$kpc, respectively.

\begin{figure}
\vspace{-0.2cm}
\epsscale{0.9}
\plotone{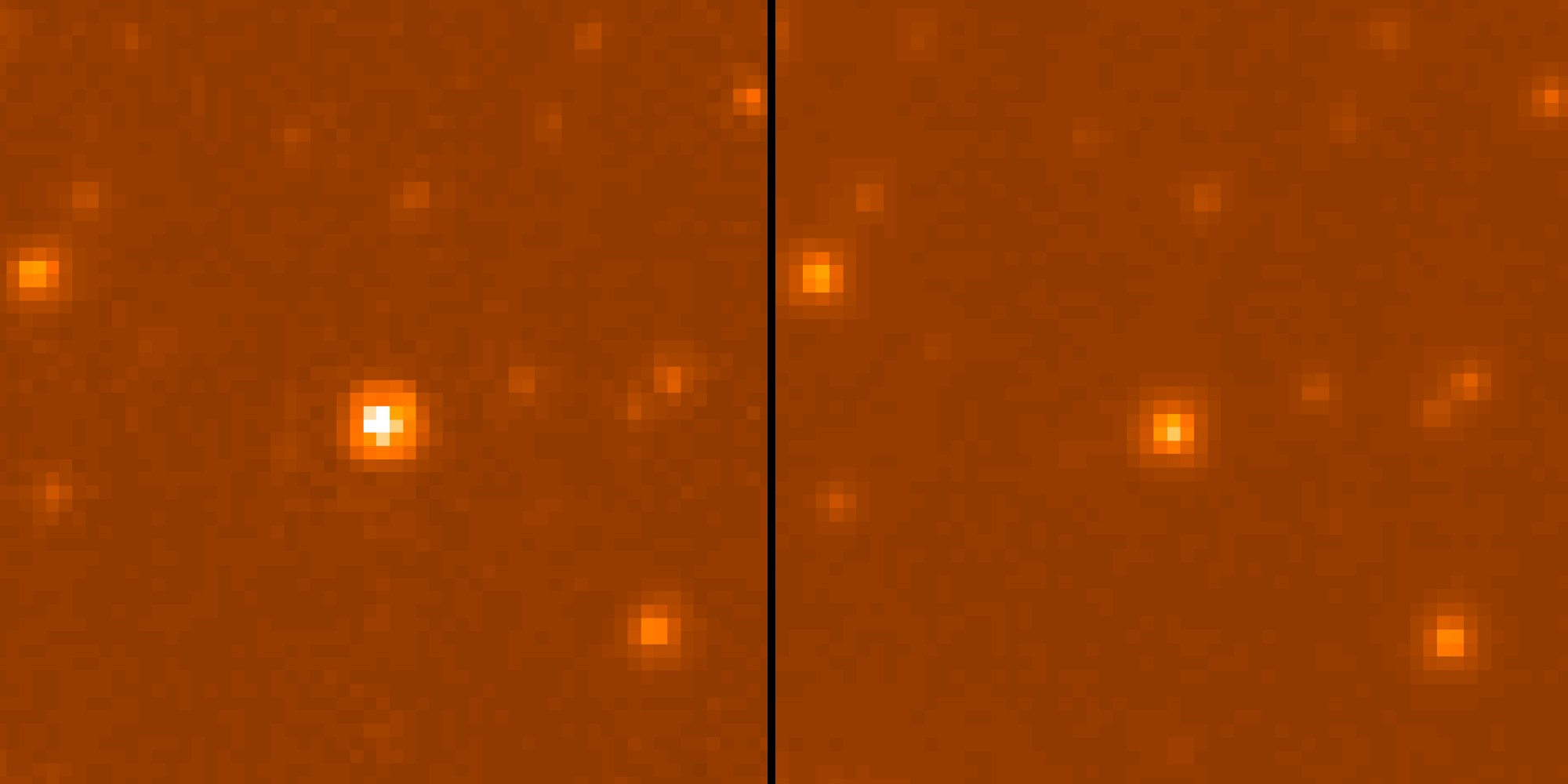}
\caption{HST WFPC2 F814W images of the OGLE-2007-BLG-349 target, while
magnified by a factor of $A = 3.444$ in the left panel and nearly
unmagnified by $A = 1.036$ in the right panel.
\label{fig-hst}}
\end{figure}

\subsection{Hubble Space Telescope Images}
\label{sec-HST}

Shortly after the planetary signal was discovered in the OGLE-2007-BLG-349
light curve, a Hubble Space Telescope Directors Discretionary proposal was submitted 
to use the Wide Field Planetary Camera 2 (WFPC2) to observe this event. This 
proposal was approved as HST program GO/DD-11352. The approved program consisted of one
short visit with a total of 320 seconds of exposures in the WFPC2 F814W passband on 8 October 2007,
some 33 days after peak magnification, as well as one longer visit on 
4 May 2008, 243 days after peak magnification. The longer visit included a total of
1280 seconds of exposures in each of the F555W and F814W passbands. The first visit
occurred when the microlensing magnification was a factor of $A = 3.444$, and the magnification
dropped to $A = 1.036$ by the time of the second observation.
Close-ups of summed images centered on the OGLE-2007-BLG-349 target from each 
visit are shown in Figure~\ref{fig-hst}, and the change in magnification of the target is clearly
visible. Because these images were taken within a year of peak magnification, the
separation between the lens and source stars \citep{bennett06,bennett07,bennett15}
is not detectable.

These HST images were reduced by two independent reduction codes. The primary reduction
used the reduction code of \citet{andking00} and \citet{andking04}, calibrated to the 
OGLE-III database \citep{ogle3-phot}, and the secondary reduction used HSTPHOT
\citep{hstphot}. The two reductions agree to better than $0.01\,$mag in absolute
calibration and better than $0.004\,$mag in the difference in magnitudes between the 
two epochs. However, these images were taken $\sim 14\,$years after the
WFPC2 instrument was installed, and the 
WFPC2 detectors have suffered significant radiation damage during this time. This
radiation damage has created defects in the detector that result in a significant
reduction in the charge transfer efficiency (CTE) of the detectors. The effect of this
CTE degradation is to reduce the sensitivity of the detectors, and we correct for this
using the tool on the Space Telescope Science Institute website 
({\tt http://www.stsci.edu/hst/wfpc2/software/wfpc2\_cte\_calc1.html}) based on the
analysis of \citet{cte_cor}. These corrections are magnitude dependent, so we have
calculated the separate corrections for the lens-plus-source target and the brighter
reference stars that are used to calibrate the HST images to the OGLE-III catalog
\citep{ogle3-phot}. For the F814W data,
the \citet{cte_cor} code indicates target magnitude corrections 
of -0.033 mag for the first epoch observations and -0.082 for the second epoch 
observations, when the event had nearly ended.
These corrected reductions give $I_{\rm HST} = 18.930 \pm 0.004\,$mag at
${\rm HJD}^\prime = 4382.0353$ and $I_{\rm HST} = 20.035 \pm 0.009\,$mag at
${\rm HJD}^\prime = 4590.7740 \equiv t_{H2}$, where ${\rm HJD}^\prime = {\rm HJD} - 2450000$.
This later measurement of $I_{\rm HST} = 20.035 \pm 0.009\,$mag, at a magnification of 
$A = 1.036$, is substantially brighter than the source magnitudes from the best fit
models presented in Table~\ref{tab-bestmparam}. It is substantially fainter than
the combined source plus lens magnitude for the best 2-planet model 
($I_{SL}(t_{H2}) = 19.162$), but it is very close to the combined source plus lens 
magnitude for the best fit circumbinary model ($I_{SL}(t_{H2}) = 20.009$). 
(The details of how the lens star magnitudes are estimated are discussed in the
next section.)
If the host star of the 2-planet model was a white dwarf, then it would
be extremely faint, and the lens plus source brightness would be just the
slightly magnified source at $I_{SL}(t_{H2}) = 20.318$, which is substantially
fainter than the $I$-band brightness measured in the HST images.
This suggests that a circumbinary model is preferred, because a 2-planet model
with a main sequence host would appear to be too bright to match the HST
data, while a 2-planet system orbiting a white dwarf would be too faint.

The WFPC2/F555W ($V$-band) images can also constrain the lens system, 
and they also support the circumbinary model. The F555W images yield a 
CTE corrected source plus lens magnitude of $V_{SL}(t_{H2}) = 22.33 \pm 0.04$. 
This compares to predictions of $V_{SL}(t_{H2}) = 21.38$ for the 2-planet
model with a main sequence host, and $V_{SL}(t_{H2}) = 22.23$ for combined 
brightness of the source and two lens stars for the circumbinary model. So,
the HST $V$-band data seem to clearly favor the circumbinary model, as well.
However,
if the planetary host star was a white dwarf, the host star brightness would be 
negligible, so it would have $V_{SL}(t_{H2}) = 22.30$. This is 
consistent with the HST $V$-band measurement.

These comparisons between the best fit 2-planet and circumbinary models
and the HST data suggest that the circumbinary model is favored, but 
to reach a firm conclusion, we need to consider more
than the best fit models. We must determine which models are consistent with both
the light curve data and the HST images. The F814W ($I$-band) images provide
a much stronger constraint than the F555W ($V$-band) images, because the
low-mass lens stars are brighter in the $I$-band and because the 
the uncertainties in both the extinction and CTE correction are larger
in the $V$-band. In the next section, we will apply a constraint from the
HST F814W observations to the light curve models, and find the light curve models
in each category that are most consistent with the light curve and HST F814W
images. We will also compare these constrained models with the HST F555W
data.

\subsection{Light Curve Models with Hubble Space Telescope Constraint}
\label{sec-HSTmod}

In order to determine which of our models are consistent with the HST imaging data,
we perform a set of constrained fits in which the lens system is forced to match the
HST observations. We consider 3 different possibilities:
\begin{enumerate}
\item 2-planet, 1-star model with a single main sequence host star.
\item 2-planet, 1-star model with a single white dwarf host star of negligible brightness.
\item 1-planet, 2-star model, with a circumbinary planet orbiting a pair of main sequence
stars.
\end{enumerate}
In principle, we could also consider circumbinary planets orbiting a binary consisting of a
at least one white dwarf, but the primary goal of this exercise is to establish that this is,
in fact, a circumbinary planet. Also, white dwarfs generally form at a late stage of stellar
evolution, after earlier stages of stellar evolution that may have removed planets from
the vicinity of the Einstein ring, where they are detectable by microlensing. (Mass loss
by stars on the giant or supergiant branch or during planetary nebula formation could
shift planets to wide orbits or unbind them from their former host star, depending on the 
details of the mass loss processes.)

At a Galactic latitude of $ b = -2.5161^\circ$, and a lens distance of $\sim 3\,$kpc, the lens system
is likely to be behind about 3/4 of the dust that is in the foreground of the source.
We model the dust with a simple exponential scale height of $h_{\rm dust} = 0.10\pm 0.02\,$kpc 
\citep{drimmel}, so that the extinction in the foreground of the lens is given by
\begin{equation}
\label{eq-extmod}
A_{i,L} = {1-e^{-|D_L/(h_{\rm dust} \sin b)|}\over 1-e^{-|D_S/(h_{\rm dust}\sin b)|}} A_{i,S} \ ,
\end{equation}
where the index $i$ refers to the passband, which is the $I$-band in this case. 

For possibility \#2, a two planet model with a white dwarf host, all the detectable flux
comes from the source star, which is directly determined by the fit. So, the uncertainty in 
the extinction plays no role. Therefore, for these models we constrain the very slightly
lensed source brightness at the time of the second epoch HST observation 
(${\rm HJD}^\prime = 4590.7740$) to be $I = 20.035 \pm 0.010$.

For possibilities \#1 and \#3, we require a mass-luminosity relation, and we use the
same empirical mass-luminosity relation that was used in \citet{bennett15}. We use the mass-luminsity
relations of \citet{henry93}, \citet{henry99} and \citet{delfosse00} in different mass
ranges. For $M_L > 0.66\,\msun$, we use the \citet{henry93} relation; for
$0.12\,\msun < M_L < 0.54\,\msun$, we use the \citet{delfosse00} relation; and for
$0.07 \,\msun < M_L < 0.10\,\msun$, we use the \citet{henry99} relation. In between these
mass ranges, we linearly interpolate between the two relations used on the
boundaries. That is we interpolate between the \citet{henry93} and the \citet{delfosse00}
relations for $0.54\,\msun < M_L < 0.66\,\msun$, and we interpolate between the
\citet{delfosse00} and \citet{henry99} relations for $0.10\,\msun < M_L < 0.12\,\msun$.

The extinction is also an important uncertainty for possibilities \#1 and \#3. We use
equation~\ref{eq-extmod} to estimate the extinction, but we also need to include a 
reasonable uncertainty for this model. About one third of the flux at the second epoch
observation (at ${\rm HJD}^\prime = 4590.7740$)
is due to the lens, so an 11\% uncertainty in the extinction in the foreground of the lens
would correspond to a 3.7\% uncertainty in the combined lens plus source flux, or a 
0.04 mag uncertainty when combined with the 0.01 mag uncertainty assumed for the
HST calibration. We therefore apply the constraint $I_{SL}(t_{H2}) = 20.035 \pm 0.040$ on the combined
source plus lens flux at ${\rm HJD}^\prime = t_{H2} = 4590.7740$.

\begin{deluxetable}{cccccccc}
\tablecaption{Best Fit Nonlinear Model Parameters with HST $I$-band Flux Constraint
                         \label{tab-constmparam} }
\tablewidth{0pt}
\tablehead{
& & & & \multicolumn{4}{c} {circumbinary} \\
& & \multicolumn{2}{c} {2-planet}  & \multicolumn{2}{c} {$u_0<0$} & \multicolumn{2}{c} {$u_0>0$} \\
\colhead{param.}  & \colhead{units} & \colhead{MS}  & \colhead{WD}  & \colhead{ $d_{1\rm cm}<1$} 
 & \colhead{$d_{1\rm cm}>1$}  & \colhead{$d_{1\rm cm}<1$}  & \colhead{ $d_{1\rm cm}>1$} 
}  

\startdata
$t_E$ & days                        & 112.758      &  95.738     & {\bf 117.793}     & 121.141     & 118.939      & 120.191 \\
$t_0$ & ${\rm HJD}^\prime$ & 4348.7472 & 4348.7467 & {\bf 4348.7465} & 4348.7520 & 4348.7459 & 4348.7511 \\
$u_0$ &                                & 0.002072   & 0.002443   & {\bf -0.001981}  & -0.002051  & 0.001966   & 0.002077 \\
$d_{1\rm cm}$ &                   & 0.79580    & 0.79647     & {\bf 0.81424}      & 1.22544     & 0.81399     & 1.22632 \\
$d_{23}$ &                            & 1.05349    & 0.95080    & {\bf 0.01903}      & 0.01951     & 0.01784     & 0.01801 \\
$\theta_{1\rm cm}$ & rad     & 1.89675     & 1.89351     & {\bf 4.35929}      & 4.35886     & 1.91859     & 1.91626 \\
$\phi_{23}$ & rad                 & -3.07236   & -3.07866    & {\bf 0.37073}     & 0.35840     & -0.40457     & -0.42858 \\
$\epsilon_1$ & $10^{-4}$      & $3.7893$ & $4.4687$    & {\bf 3.4119}        & 3.3134       & 3.3895        & 3.3471 \\
$\epsilon_2$ & & $9.304\times 10^{-6}$  & $9.827\times 10^{-6}$ & {\bf 0.45967} & 0.42723 & 0.48342 & 0.48489 \\
$\epsilon_3$ &                      & 0.99961   & 0.99954       & {\bf 0.53999}    & 0.57244     & 0.51624      & 0.51477 \\
$t_\ast$ & days                     & 0.06954    & 0.06929      & {\bf 0.07048}    & 0.07072     & 0.07074      & 0.07056 \\
$\dot{d}_{23x}$ & ${\rm days}^{-1}$ & 0.0 & 0.0             & {\bf 0.010586}  & 0.012073    & 0.008882   & 0.009545 \\
$\dot{d}_{23y}$ & ${\rm days}^{-1}$ & 0.0 & 0.0             & {\bf -0.006420} & -0.006506  & 0.005603   & 0.003518 \\
$1/T_{\rm orb}$ & ${\rm days}^{-1}$ & 0.0 & 0.0             & {\bf 0.060459} & 0.070847     & 0.059174   & 0.059632 \\
$\pi_E$ &                              & 0.36051     & 0.19544     & {\bf 0.18361}   & 0.20926       & 0.13819 & 0.14066 \\
$\phi_E$ & rad                      & 2.81614     & 2.38046    & {\bf 0.59559}   & 0.48368       & 0.79651    & 0.76876 \\
$\theta_E$ & mas                 & 1.1082       & 1.0261       & {\bf 1.1167}     & 1.1282         &  1.1184       & 1.1273 \\
$M_L$ & $\msun$                & 0.3775       & 0.6447        & {\bf 0.7468}    & 0.6620         &  0.9937   & 0.9841 \\
$I_{Sc}$ &  mag                   & 20.298       & 20.138       & {\bf 20.365}    & 20.396         & 20.375      & 20.386 \\
$I_{Sh}$ &  mag                   & 20.317       & 20.072       & {\bf 20.371}    & 20.409         & 20.389      & 20.403 \\
$I_L$ &  mag                        & 21.006       & -                 & {\bf 21.527}   & 21.484          & 21.410      & 21.390 \\
$I_{SL}(t_{H2})$ & mag        & 19.826       & 20.053       & {\bf 20.020}    & 20.036         & 20.001       & 20.005 \\
$V_S$ & mag                       & 22.328       & 22.148       & {\bf 22.375}   & 22.406         & 22.385      &  22.396 \\
$H_S$ & mag                       & 18.219       & 18.040       & {\bf 18.268}   & 18.267         & 18.277      &  18.288 \\
fit $\chi^2$ &                         & 3438.88     & 3426.19    & {\bf 3382.43}  & 3387.85       & 3408.17 & 3412.62 \\
dof &                                     & 3569          & 3569         & {\bf 3567}       & 3567            & 3567       & 3567 \\
\enddata
\tablecomments{${\rm HJD}^\prime = {\rm HJD}-2$,450,000. The best fit parameters are indicated in 
bold face.}

\end{deluxetable}

For the circumbinary models, the fit parameters (along with $\theta_*$) determine the 
source distance, if we insist that the stellar orbits be circular. (The modeling employs circular 
orbits, but these can be interpreted as second order approximations to any bound orbit.
See Bennett \etal\ 2010 for more discussion of this point.) With $\theta_E$ and $\pi_E$
determined, we know the mass of the lens system, via equation~\ref{eq-M}, and we also know the 
5 parameters describing the orbit, $d_{23}$, $\phi_{23}$, $\dot{d}_{23x}$, $\dot{d}_{23y}$,
and $1/T_{\rm orb}$ in Einstein radius units. We only need the distance to the lens system
to convert these to physical units, and this is given by equation~\ref{eq-Dl} (assuming that we
already know $D_S$). However, we already
know the size of the orbit in physical units, via Kepler's third law, since we know the period
and the mass of the binary host system. 
So, we can use this information to invert equation~\ref{eq-Dl} and solve
for the source distance. From the CMD in Figure~\ref{fig-cmd}, we see that the source lies on the 
red side of the bulge main sequence, and we know that the red color
is explained by the high metallicity measured by \citet{bensby13}. So, it is safe to assume that
the source is located in the bulge. We therefore apply a constraint on the implied
distance to the source in the circumbinary models, $D_S = 7.8\pm 1.4\,$kpc, assuming
the bulge distance estimate by \citet{nataf13} at the Galactic longitude of this event.

Table~\ref{tab-constmparam} gives the parameters of the best fit models with the source plus 
lens $I$-band magnitude ($I_{SL}$) constraint imposed at the time, $t_{H2}$, of the second
epoch of HST observations. The best fit circumbinary model does very well with the 
constraint on the $I$-band lens-plus-source brightness, as the application of this constraint
only increases $\chi^2$ by $\Delta\chi^2 = 0.18$ for one additional degree of freedom.
The two-planet models are so significantly disfavored that we can
exclude them based on these constrained fits. The best two-planet model with a main sequence host
is disfavored by $\Delta\chi^2 = 56.45$ with respect to the best circumbinary model, and the 
two-planet model with a dark stellar remnant
host is disfavored by $\Delta\chi^2 = 43.76$. For the main sequence host case, 
$\Delta\chi^2 = 33.13$ comes from the $I_{SL}(t_{H2})$ constraint and $\Delta\chi^2 = 23.32$
comes from the difference in the light curve model fits. In the case of a dark stellar 
remnant host, almost the entire $\Delta\chi^2$ difference comes from 
the light curve difference. These $\chi^2$ differences
are sufficient to exclude both the two-planet and white dwarf host models. If we assume 
Gaussian random errors, then the probability of the best non-circumbinary solution is
$3\times 10^{-9}$. A very conservative choice would be to substitute $\Delta\chi^2/2$ 
for $\Delta\chi^2$ into the $\chi^2$ probability distribution formula. This is equivalent to
to assuming that the correlations and non-Gaussianity of the errors have the same
effect as increasing each error bar (and constraint) by a factor of $\sqrt{2}$.
With this assumption, the probability of the best non-circumbinary model would be 
$2\times 10^{-5}$, so even with a very conservative assumption about the effects of
non-Gaussian and correlated errors, it is only the circumbinary planetary models that
are viable. We also note that the circumbinary models can be confirmed by observing
the lens stars with the predicted brightness of $I_L = 21.39\pm 0.24$ and
$H_L = 18.57\pm 0.22$ separating 
from the source at the predicted rate of $\mu_{\rm rel} = 3.55\pm 0.15\,$mas/yr.
(These numbers come from the MCMC calculations discussed later in this section.)

We can also compare the measured $V$-band HST (F555W) magnitudes to the
predicted values from these constrained models. The best fit constrained 
circumbinary model predicts $V_{SL}(t_{H2}) = 22.26 \pm 0.07$, which
compares to the measured value of $V_{SL}(t_{H2}) = 22.33 \pm 0.04$
for $\Delta\chi^2 = 0.75$. The 2-planet models don't do as well. With a main
sequence host star, we have the prediction of $V_{SL}(t_{H2}) = 22.11 \pm 0.07$,
which is still too bright and implies $\Delta\chi^2 = 7.44$. For a white dwarf host,
which matched the HST $V$-band measurement without the HST $I$-band
constraint, the constraint has pushed the $V$-band magnitude to be too
bright, $V_{SL}(t_{H2}) = 22.13 \pm 0.01$. Comparison to the measurement,
yields $\Delta\chi^2 = 23.53$, so it is only the circumbinary model that is consistent
with the HST $V$-band measurement.

This comparison with HST images also rules out the models
in which the planet orbits one member of a wide binary star system, although
these were already excluded due to the lack of an acceptable light curve
fit. The microlensing parallax measurement constrains the mass interior
to the Einstein radius of the primary lens mass. This is the single host
star of the two planet models or both host stars for a circumbinary system.
For the case of a planet orbiting one star of a wide binary system, the 
mass constrained by the microlensing parallax measurement is the 
mass of the planetary host star. Its wide binary companion just provides
a small perturbation to the light curve. So, these models are also excluded
by the same argument that excludes the 2-planet models. The microlensing
parallax measurement requires a planetary host mass of $\sim 0.7\msun$,
and it is only if the host is a close binary system that this mass can be
split into two stars that are consistent with the HST images.

\begin{deluxetable}{ccc}
\tablecaption{Constrained Circumbinary MCMC Model Parameters
                         \label{tab-mcmcparam} }
\tablewidth{0pt}
\tablehead{
\colhead{parameter}  & \colhead{units} & \colhead{Mean MCMC Values} 
}  

\startdata

$t_E$ & days & $118\pm 4$  \\
$t_0$ & ${\rm HJD}^\prime$ & $4348.7470\pm 0.0014$  \\
$u_0$ & & $-0.00198\pm 0.00007$  \\
$d_{1\rm cm}$ & close mod. & $0.8146\pm 0.0015$  \\
($d_{1\rm cm}$) & wide mod. & $1.2257\pm 0.0022$  \\
$d_{23}$ & & $0.0193\pm 0.0010$  \\
$\theta_{1\rm cm}$ & radians & $4.3590\pm 0.0030$  \\
$\phi_{23}$ & radians & $0.368\pm 0.056$  \\
$\epsilon_1$ & & $(3.39\pm 0.12)\times 10^{-4}$ \\
$\epsilon_2$ & & $0.493\pm 0.098$ \\
$\epsilon_3$ & &  $0.507\pm 0.098$ \\
$t_\ast$ & days &  $0.07065\pm 0.00043$ \\
$\dot{d}_{12x}$ & ${\rm days}^{-1}$ & $0.0096\pm 0.0023$ \\
$\dot{d}_{12y}$ & ${\rm days}^{-1}$ & $-0.0069\pm 0.0043$ \\
$1/T_{\rm orb}$ & ${\rm days}^{-1}$ &  $0.061\pm 0.091$\\
$\pi_E$ & & $0.204\pm 0.034$ \\
$\phi_E$ & radians &  $0.54\pm 0.11$ \\
$\mu_{\rm rel}$ & mas/yr & $3.55\pm 0.15$ \\
$\theta_E$ & mas & $1.15 \pm 0.05$ \\
$V_S$ & mag & $22.380 \pm 0.037$ \\
$I_S$ & mag & $20.369 \pm 0.037$ \\
$H_S$ & mag & $18.272 \pm 0.037$ \\
\enddata

\tablecomments{ ${\rm HJD}^\prime = {\rm HJD}-2$,450,000. The close model is preferred
over the wide model by $\Delta\chi^2 = 4.92$.
}

\end{deluxetable}

For the circumbinary models, we present the best fit model parameters for each of the degenerate 
solutions, with $u_0<0$ or $u_0>0$ and with $d_{1\rm cm}<1$ or $d_{1\rm cm}>1$,
in Table~\ref{tab-constmparam}.
The $u_0>0$ models have smaller parallaxes and therefore larger host star masses,
and this means that they are disfavored by the $I_{SL}(t_{H2})$ constraint by
$\Delta\chi^2 \sim 26$-30. The wide models
with $d_{1\rm cm} \approx 1.225$ are also slightly disfavored by $\Delta\chi^2 \sim 5$.

\begin{deluxetable}{cccc}
\tablecaption{Physical Parameters\label{tab-pparam}}
\tablewidth{0pt}
\tablehead{
\colhead{Parameter}  & \colhead{units} & \colhead{average value} & \colhead{2-$\sigma$ range} }
\startdata 
$D_L $ & kpc & $2.76\pm 0.38$ & 2.06-3.56 \\
$M_{\rm A+B}$ & $\msun$ & $0.71 \pm 0.12$ & 0.50-0.95 \\
$M_{\rm A}$ & $\msun$ & $0.41\pm 0.07$ & 0.28-0.54 \\
$M_{\rm B}$ & $\msun$ & $0.30 \pm 0.07$ & 0.15-0.45 \\
$m_c$ & $\mearth$ & $80\pm 13 $ & 56-107 \\
$a_{\perp \rm AB}$ & AU & $0.061 \pm 0.007$ & 0.048-0.074  \\
$a_{\rm AB}$ & AU & $0.080{+0.027\atop -0.015}$ & 0.054-0.162  \\
$P_{\rm AB}$ & days & $9.7{+5.4\atop -2.5}$ & 5.7-28.1 \\
$a_{\perp {\rm CM}c}$ & AU & $2.59 {+0.43\atop -0.34}$ & 1.97-3.89  \\
$V_L$ & mag & $24.73\pm 0.34$ & 24.18-25.51 \\
$I_L$ & mag & $21.39\pm 0.24$ & 20.98-21.98 \\
$H_L$ & mag & $18.57\pm 0.22$ & 18.22-19.07 \\
\enddata
\tablecomments{The average value is the mean value for all parameters except that we 
use the median for $a_{\perp \rm AB}$,
$a_{\rm AB}$, $P_{\rm AB}$, and $a_{\perp {\rm CM}c}$. }
\end{deluxetable}

In order to determine the ranges of parameters and properties that are consistent
with the observed light curve and HST constraint, we have performed a 
series of Markov Chain Monte Carlo (MCMC) \citep{verde03} runs. We follow the
usual procedure 
\citep{bennett08} of weighting each class of models with the weight function, $e^{-\Delta\chi^2/2}$,
where $\Delta\chi^2$ refers to the $\chi^2$ difference between the local $\chi^2$ minimum and
the global $\chi^2$ minimum (with $u_0 < 0$ and $d_{1\rm cm} < 1$). For the 
$u_0 < 0$, $d_{1\rm cm} > 1$ models, the penalty is $\Delta\chi^2 = 5.42$, which corresponds
to a weight of 0.067, but for the $u_0 > 0$ models the penalties are $\Delta\chi^2 = 25.74$ 
(for $d_{1\rm cm} < 1$) and $\Delta\chi^2 = 30.19$ (for $d_{1\rm cm} > 1$), 
corresponding to weights of $3\times 10^{-6}$ and $3\times 10^{-7}$, respectively.
The weights for the $u_0 > 0$ models are so small that they don't contribute to the 
mean microlens model parameters, shown in Table~\ref{tab-mcmcparam}. The physical
parameters of the lens system from these MCMC runs are given in Table~\ref{tab-pparam}.
The system consists of a planet of $80\pm 13\mearth$ orbiting binary stellar system,
consisting of two M-dwarfs with masses of $0.41 \pm 0.07\msun$ and $0.30\pm 0.07\msun$.
These stars have a semi-major axis of $0.080{+0.027\atop -0.015}\,$AU and a period of
$9.7{+5.4\atop -2.5}\,$days. The median two-dimensional separation of the planet from the stellar
center-of-mass is $2.59{+0.43\atop -0.34}\,$AU, which implies a median semi-major axis of
$\sim 3.2\,$AU and an orbital period of about $7\,$years if we assume a random orbital
orientation. However, it is only the transverse separation that is measured (and reported in
Table~\ref{tab-pparam}), so the 3-dimensional separation could be much larger
if the line-of-sight separation between the planet and binary stars is large.
Then, the semi-major 
axis and orbital period could be substantially larger than this.

\section{VLT/NACO Observations of the Source Plus Lens System}
\label{sec-NACO}

We have also obtained two epochs of adaptive optics observations in the
infrared $JHK$ passbands with the Very Large Telescope (VLT)
NACO instrument with a $28^{\prime\prime}\times 28^{\prime\prime}$ field-of-view (FOV).
The first $H$-band observations were taken at ${\rm HJD}^\prime = 4386.046880$, 
when the magnification was about a factor of 3, and the second epoch observations
were taken at ${\rm HJD}^\prime = 4686.144531$, when the magnification was only
about 1\%.
This small field-of-view made calibration of the VLT/NACO data very difficult, and, in
fact, we were unable to find a satisfactory calibration of these data. We were able
to calibrate the CTIO H-band data as discussed in Section~\ref{sec-radius}, and so from the 
CTIO $H$-band light curve, we know the $H$-band source magnitude. 

Each of the different models presented in Table~\ref{tab-constmparam} has a different 
prediction for the combined lens plus source magnitude at the times of the two different 
observations. The observations give a magnitude difference of $0.87\pm 0.10$ between
the two epochs. This compares to the 0.575, 1.010, and 0.680 source plus
lens $H$-band magnitude difference between the first and second epochs for the 
2-planet plus main sequence host, 2-planet plus stellar remnant
host, and circumbinary models, respectively. These differ from the measured value by 
2.95-$\sigma$, 1.40-$\sigma$, and 1.90-$\sigma$, respectively. So, the VLT/NACO data slightly favor
the circumbinary model over the 2-planet model with a main sequence host, but 
a white dwarf host does slightly better. The effect is too small to alter our 
conclusions, however.

\section{Discussion and Conclusions}
\label{sec-conclude}

In the previous section, we have established that although the OGLE-2007-BLG-349 light curve
can be explained by models with one star and two planets, it is only the circumbinary planet
models that can explain both the light curve and the HST observations. So, the system consists of two host stars,
OGLE-2007-BLG-349LA and OGLE-2007-BLG-349LB, orbited by a planet somewhat 
less massive than Saturn. Although it was the first circumbinary planet to be observed,
aside from a planet orbiting a neutron star-white dwarf system \citep{ford00,sig03}, it
was not the first circumbinary planet to be published, as 10 circumbinary planets
\citep{doyle11,welch15,kostov16} have been discovered by the Kepler mission.

\begin{figure}
\vspace{-0.2cm}
\epsscale{0.8}
\plotone{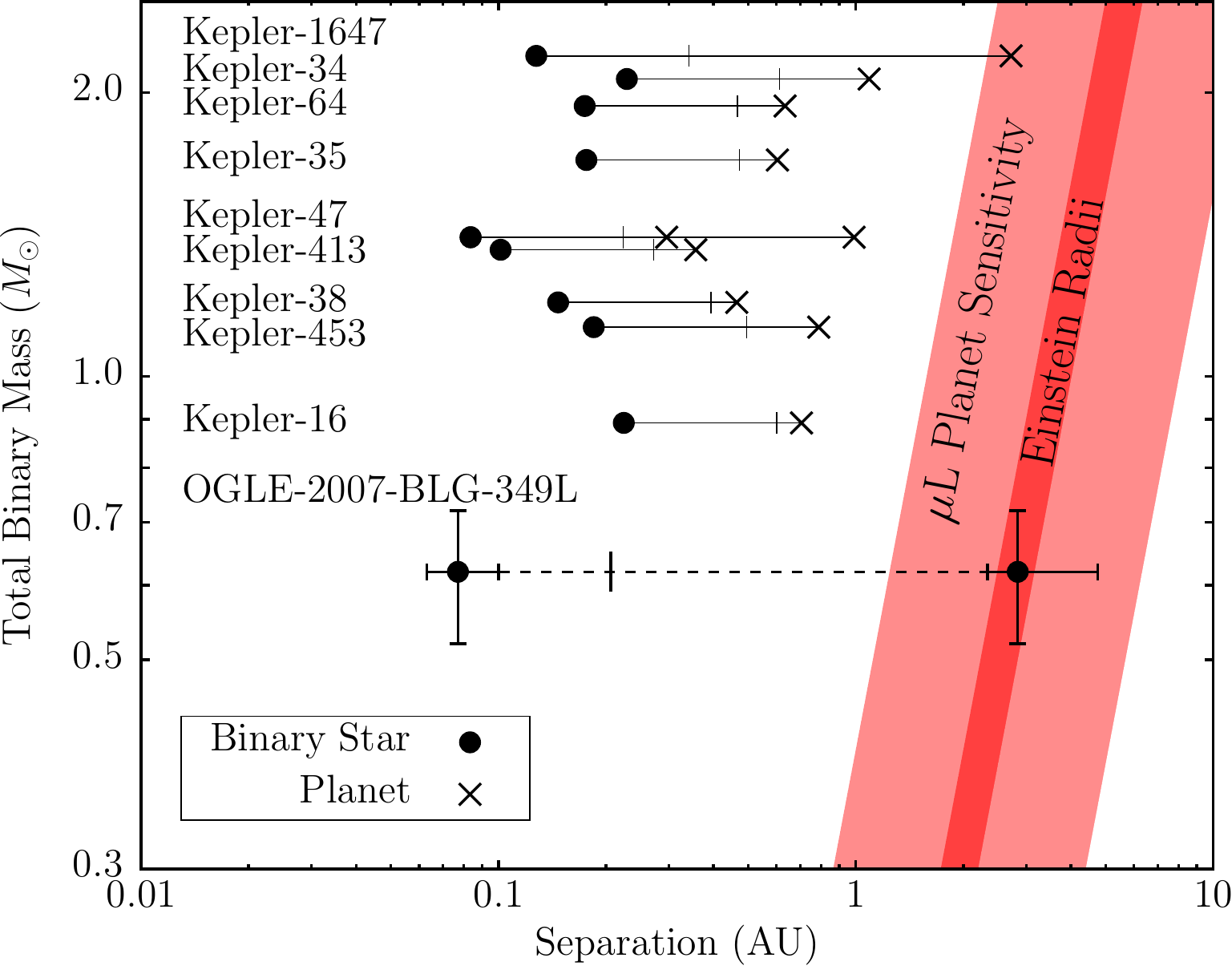}
\caption{Comparison of host star masses and orbital separations for the
known circumbinary planet systems. The filled circles show the orbital
separations of the host stars, while the orbital separations of the planets from
the stellar centers of mass are marked with ``x"s. The vertical bars on each line
indicate the approximate stability limit. The red region gives the typical
Einstein radius as a function of mass and the light red region gives the approximate
range of planetary microlens sensitivity.
\label{fig-circumbin}}
\end{figure}

One puzzle with the circumbinary planets discovered in the Kepler data is that most
of them are located quite close to the stability limit \citep{holman99}, as shown in 
Figure~\ref{fig-circumbin}. That is, if they
were moved to orbits with slightly smaller semi-major axes, they would quickly become
dynamically unstable. \citet{holman99} find that circular,
coplanar circumbinary orbits become unstable within
$a_c \simeq (2.28\pm 0.01) + (3.8\pm 0.3)e + (1.7\pm 0.1)e^2$, where $e$ is the 
eccentricity of the binary orbit and $a_c$ is measured in units of the stellar binary
semi-major axis. Our modeling has enforced a circular orbit for the stellar binary,
so it is sensible to assume a low eccentricity. Also, if the stellar binary orbit does
have a significant eccentricity, then it is likely that the semi-major axis is smaller
than the mean values listed in Table~\ref{tab-pparam}, because stars spend most of their
time in an eccentric orbit at separations larger than the semi-major axis. So, the consideration
of stellar binary orbits with significant eccentricity is not likely to significantly increase the
maximum stellar separation, which is closely related to the stability constraint.
So, we assume $e \approx 0.1$, and this yields $a_c = 2.7$. The median
semi-major axis of the OGLE-2007-BLG-349LAB binary is $a \simeq 0.080\,$AU 
from Table~\ref{tab-pparam}, and the median three-dimensional separation between 
stellar center-of-mass and the OGLE-2007-BLG-349L(AB)c planet is $\sim 3.2\,$AU.
Therefore, we estimate the planet orbits at $\sim 15a_c$. This compares to most of the
Kepler circumbinary planets that orbit at $< 2a_c$, and the widest orbit Kepler
circumbinary planet \citep{kostov16} that orbits at $7a_c$.

The expected orbital period for the OGLE-2007-BLG-349L(AB)c planet is
$\sim 7\,$years assuming a host system mass of $0.71\,\msun$ and a semi-major
axis of $3.2\,$AU, so such a system could not have been detected by Kepler.
The only Kepler planet with a comparable separation is Kepler-1647b. It
orbits a star system that is three times more massive than the OGLE-2007-BLG-349L
host star system, which implies a period short enough to allow for its detection
with two transit episodes during Kepler observations. We expect that Kepler's detection
efficiency for such systems is quite low, so such systems might be quite common.

The fact that the first circumbinary planet found by microlensing has an orbital separation
well beyond the stability limit adds modest support to the idea that circumbinary
planets far beyond the stability limit are quite common. This would imply that
circumbinary planets probably form in the outer disk, relatively far from the orbital stability
limit \citep{kley14,bromley15,silsbee15} instead of in situ \citep{meschiari14}. In principle,
this new microlensing discovery could provide strong evidence that circumbinary 
planets are substantially more common far from the stability limit than close to the 
stability limit \citep{luhn16}. Microlensing is most sensitive to both planets and stellar companions
at separations close to the Einstein radius.
However, for event OGLE-2007-BLG-349,
the ratio of the two-dimensional separation between the planet and stellar center-of-mass
to the separation between the two stars is 42. Such a large ratio was only detectable because of
the very high magnification of this event, but circumbinary planets with a smaller
separation ratio should be detectable for a much larger class of lower-magnification
events. The fact that no other circumbinary planets have been found by microlensing
might be considered to imply that circumbinary planets with smaller separation ratios are
more rare. However, there is circumstantial evidence suggesting that we may be 
inefficient at identifying such events in our data. Unlike the transit method,
microlensing is sensitive to planets beyond the snow line,
\citep{lecar_snowline,kennedy_snowline}, so OGLE-2007-BLG-349L(AB)c is the 
first circumbinary planet beyond the snow line.

\citet{gould14} presented another two-star plus one planet event, OGLE-2013-BLG-0341,
which was interpreted as a wide binary with a planet orbiting one of the two stars,
although there are circumbinary models with very similar light curves. This was also a 
high magnification event with the signal dominated by the stellar binary instead of by
the planet (like OGLE-2007-BLG-349). However, the lens-source alignment was such
that the source crossed a planetary caustic feature prior to reaching high magnification.
This made it obvious that the lens system included a planet, but we were very lucky 
to have this planetary feature detected. And the analysis showed that the planet was
required to fit the data even if the low-magnification planetary feature was not seen.
This suggests that there should be many more two-star plus one planet events in
the data that we have already collected, but that we are not efficient at finding
planetary signals in events that are dominated by stellar binary microlensing
features. Hence, we recommend a systematic search for planetary signals in the light
curves of strong stellar binary events. If a large population of circumbinary planets
are found, it will add to the $\sim 10$\% frequency of circumbinary planets found
in short period orbits \citep{armstrong14}. Circumbinary planetary systems can be
quite efficient at ejecting planets \citep{sutherland16,smullen16}, so they could contribute to
the large population of rogue planets found by microlensing \citep{sumi11}.

\acknowledgments 
The authors would like to thank Subo Dong for a great deal of work on this
event, including the initial realization that a 3rd mass was needed to fit the
light curve.
D.P.B., A.B., and D.S.\  were supported by NASA through grants NASA-NNX12AF54G
and NNX13AF64G. 
The MOA group acknowledges financial support from the Marsden Fund of NZ
and in Japan from grants JSPS25103508 and 23340064.
T.S. received support from JSPS23103002, JSPS24253004 and 
JSPS26247023. A.G.\ and B.S.G.\ were supported by NSF grant AST-1516842.
OGLE Team thanks Profs.\ M. Kubiak and G. Pietrzy{\'n}ski, former
members of the OGLE team, for their contribution to the collection of
the OGLE photometric data over the past years.
The OGLE project has received funding from the National Science Centre,
Poland, grant MAESTRO 2014/14/A/ST9/00121 to AU.
NASA/ESA Hubble Space Telescope data from program GO/DD-11352 was obtained from the 
data archive at the Space Telescope Science Institute. STScI is operated by the
 Association of Universities for Research in Astronomy, Inc. under NASA contract NAS 5-26555.
 Work  by  C.  Han  was  supported  by  the  Creative  Research Initiative  Program
(2009-0081561)  of  National  Research  Foundation  of  Korea.
 KH acknowledges support from STFC grant ST/M001296/1.






\end{document}